Review

# Aromatic Species in the Molecular Universe

*Published as part of ACS Earth and Space Chemistry special issue "Eric Herbst Festschrift".*

Alexander G. G. M. Tielens*

ACCESS | 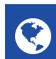 Metrics & More | Article Recommendations |

**ABSTRACT:** Interstellar polycyclic aromatic hydrocarbon (PAHs) are an important component of the interstellar medium of galaxies, containing some 10% of the elemental carbon. Their vibrational emission dominates the mid-infrared spectra of galactic and extragalactic objects. PAHs control the heating of interstellar neutral gas and the charge balance of molecular clouds. PAHs are formed in the outflows from late type stars through chemical processes akin to those in sooting flames and then further processed in the interstellar medium by UV photolysis and strong shock waves. PAHs are also formed through ion−molecule reactions and neutral−radical reactions in dense cloud cores. The James Webb Space Telescope has provided a wealth of high-quality spectra that have provided new insights in the characteristics of the interstellar PAH family. Their analysis is supported by dedicated laboratory and quantum chemistry studies, feeding into detailed molecular physics models relevant to astronomical environments. Laboratory studies have also provided deeper insight in the chemical evolution of PAHs in the interstellar medium. This paper will review progress in the field and chart its future.

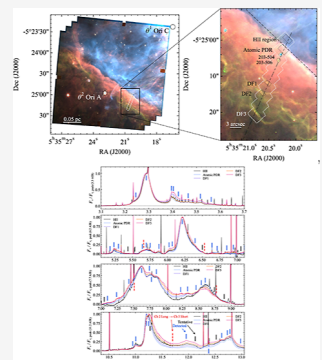

**KEYWORDS:** *Polycyclic Aromatic Hydrocarbons (PAHs), Aromatic Infrared Bands (AIBs), ), Diffuse Interstellar Bands (DIBs), interstellar medium, infrared spectroscopy, photophysics, photochemistry, ion-molecule reactions, neutral-radical reactions*

## 1. INTRODUCTION

In the mid-1970ies, ground-based and airborne spectroscopy revealed the presence of broad emission features at 3.3, 6.2, 7.7, 8.6, and 11.3 $\mu m$—the so-called Aromatic Infrared Bands (AIBs)—in the spectra of objects illuminated by strong UV sources such as H II regions and planetary nebulae. The Infrared Space Observatory (ISO) and the Spitzer Space Telescope—launched in 1995 and 2003, respectively—unambiguously demonstrated that the AIBs dominate the mid-IR spectra of C-rich post-AGB objects and planetary nebulae, the photodissociation regions (PDRs) associated with regions of massive star formation—including H II regions powered by O stars and reflection nebulae illuminated by late B stars—planetary disks associated with young stars (such as Herbig AeBe stars and T Tauri stars), the general interstellar medium of galaxies, starburst regions associated with galactic nuclei and many UltraLuminous InfraRed Galaxies (c.f., Tielens[1] and references therein). The launch of the James Webb Space Telescope (JWST) in 2021 has provided astronomers with high spatial (0.1−0.3″, depending on wavelength) and spectral resolution ($R \simeq 2500$) spectrometers covering the 0.6−28 $\mu m$ wavelength range. These very sensitive instruments promise to revolutionize our view of the AIBs and the first results bear this out.

In the mid-1980ies, the AIBs were linked to vibrational fluorescence of large Polycyclic Aromatic Hydrocarbon (PAHs) molecules containing some 30−100 C atoms.[2,3] PAHs are ubiquitous in the interstellar medium of galaxies and are very abundant—locking up some 10% of the elemental carbon. They dominate the mid-IR emission of galaxies and play an important role in the energy balance of interstellar gas and in the ionization balance of dense molecular cloud cores.[1] Observations reveal subtle variations in the detailed characteristics of the AIBs within sources and between sources and this has been ascribed to variations in the emitting PAH family linked to the local physical conditions in space. Astronomers use these variations as a tool to study variations in the properties of space.

In the early 2010s, the fullerene, $C_{60}$, was unambiguous detected in space through its vibrational and electronic signatures.[4,5] The presence of this large molecule has been attributed to photochemically driven, top-down chemistry from large PAHs.[6] While the AIBs clearly demonstrated the importance of large PAHs in space, identification of specific PAH molecules took until the early 2020ies when deep surveys in the millimeter regime using the Greenbank and Yebes telescopes revealed the rotational signature of simple aromatic molecules in dense molecular cloud cores.[7,8] In recent years, the identification of cyanopyrene and cyanocoronene in the TMC1 molecular cloud core provides a clear link to the PAH carriers of the AIBs.[9,10] This suggests a bottom-up formation route inside cold, dense cores for PAHs starting from small hydrocarbon radicals. Also, rather than using generic models, the identification of specific molecules in space (C₆₀, naphthalene, pyrene, coronene) will enable detailed model for their response to the unique conditions in space using the specific, laboratory measured or quantum chemically calculated, molecular properties of these species.

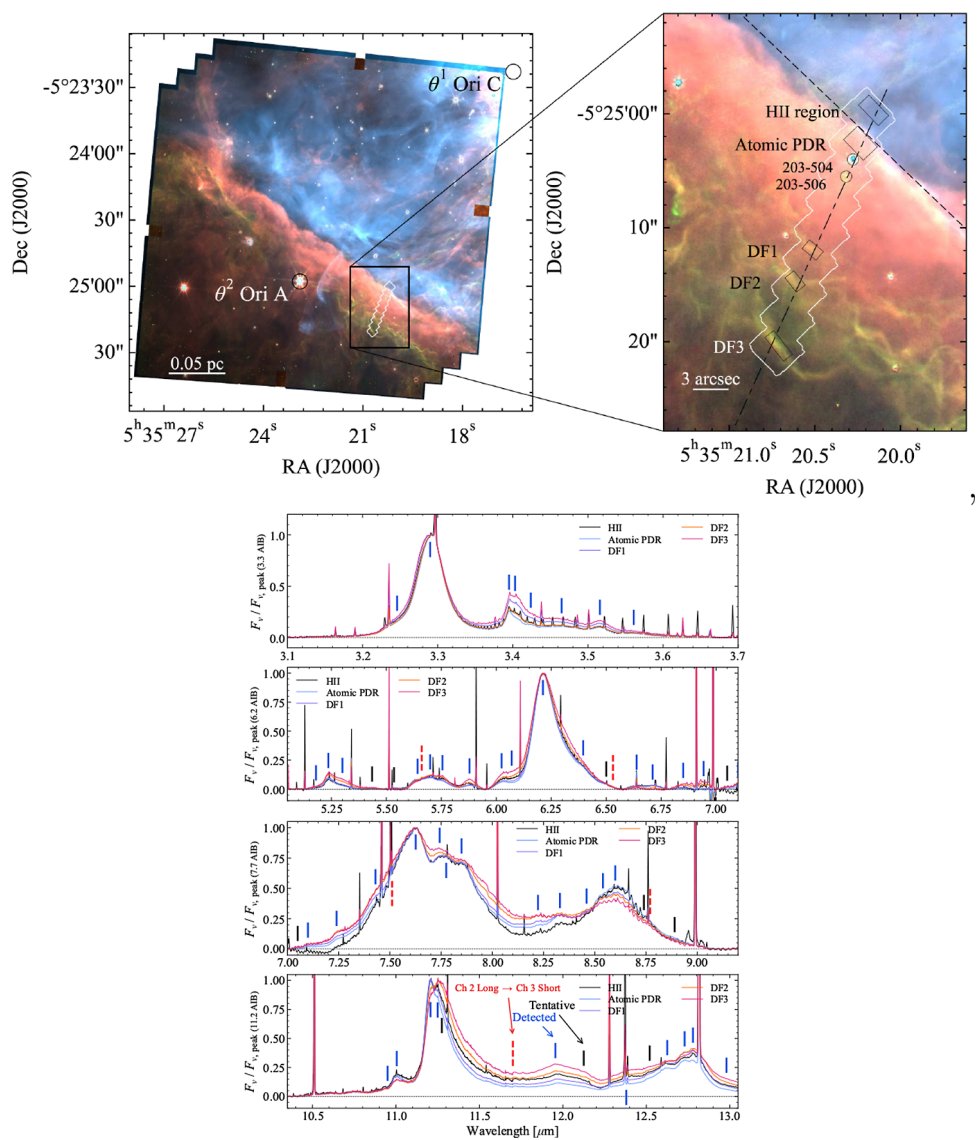

**Figure 1.** Top: Left: false color NIRCam image of the Orion Bar[11] showing the NIRSpec mosaic footprint (white boundary). The composite image combines AIB emission (red), H₂ emission (green), and H I Paschen α (blue). Top Right: Blow up of the area mapped spectrally with NIRSpec and MIRI. The black boxes define the apertures where the spectra shown in the bottom panel were extracted.[12] DF1−3 indicate the positions of the H₂ dissociation fronts. The location of two proplyds in the Orion Bar are marked by small circles. Bottom: the incredibly rich, Aromatic Infrared Band (AIB) spectrum as observed in the Orion Bar.[13] A continuum was subtracted to highlight the AIBs. The different colored spectra refer to templates for five distinct regions in the Orion PDR, the (background PDR behind the) H II region, the atomic zone in the PDR, and the three dissociation fronts that traverse the region spectrally imaged. Each spectral window (on an $F_\nu$ scale) is normalized by the peak surface brightness of the indicated AIB on the y-axes in each panel. The vertical tick marks indicate the positions of identified (blue) or tentative (black) AIBs and components. Red dashed vertical ticks indicate the wavelengths where the data switches from one MIRI subband to the next. Figure taken from.[13] Reproduced with permission from Astronomy & Astrophysics, © ESO.

Over the last 30 years, in a concerted effort, spectroscopists, molecular physicists, and physical chemists have investigated the vibrational properties and the formation and fragmentation of PAHs under conditions relevant to space and these have been included in astronomical models (c.f., Tielens[1] and references therein). The low densities and low UV fields make space a unique environment where large molecules can survive for hundreds of millions of years. PAHs are also abundant in carbonaceous meteorites and may therefore have contributed to the organic inventory of the early Earth and interstellar chemistry may have contributed to the abiotic origin of life on Earth and, potentially, other planets and moons in the solar system and exoplanets around their stars.

These exciting developments make a new review of the field very timely. Section 2 will review recent observational highlights obtained with JWST. Progress in the area of the intrinsic spectroscopic characteristics of PAHs are summarized in Section

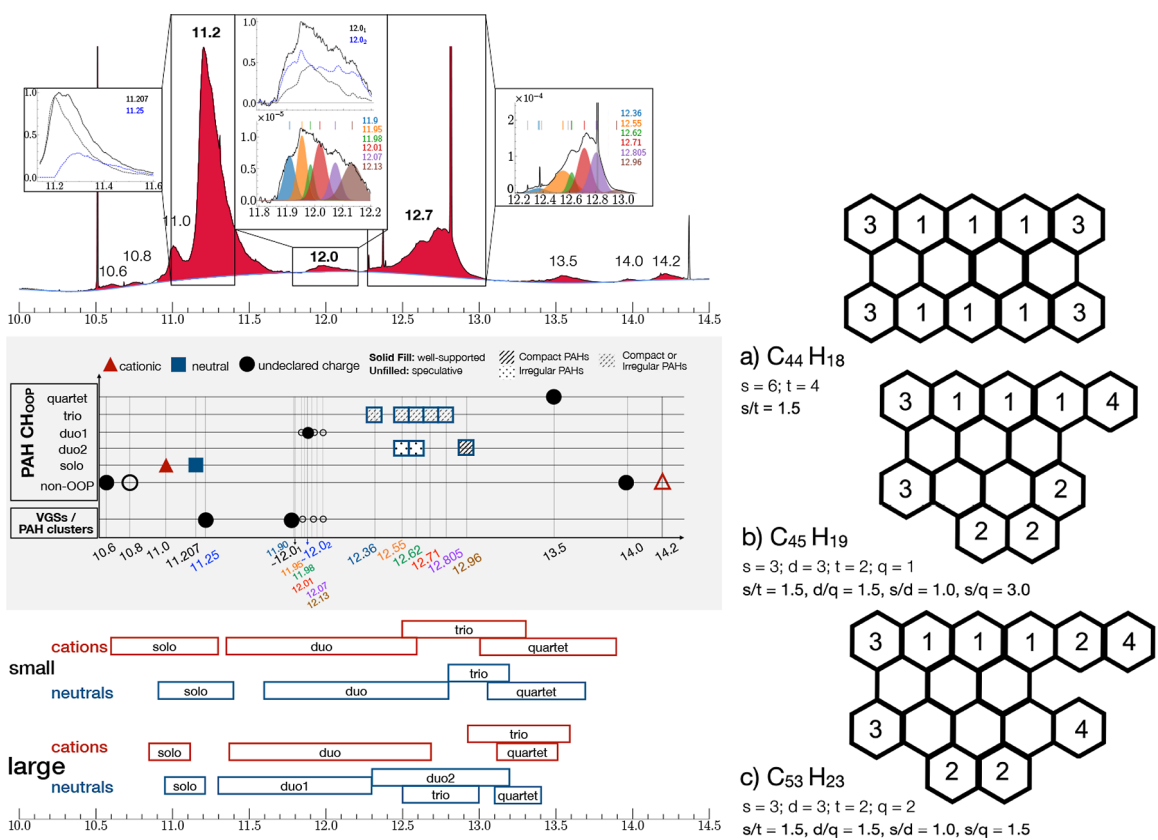

**Figure 2.** Left: analysis of the out-of-plane (OOPs) bending mode region of the Orion Bar spectra.[18] See text for details. Top panel: decomposition of the OOPs modes in the Orion Bar spectra in separate components. The colored profiles in the insets provide the individual components in the 11.2, 12.0, and 12.7 μm AIBs. These components have been extracted through spatial-spectral analysis of the Orion Bar spectra. Bottom panel: DFT quantum chemistry calculations have revealed the range—indicated by colored blocks—in which different CH edge-structures (solo, duos, trios, and quartets) emit. These ranges are split out into those for small and large, cationic and neutral PAHs Middle panel: Assignment of the identified spectral components in the AIBs to the different cationic and neutral PAHs. Right: representative molecular structures characteristic for the edge structure in the interstellar PAH family as derived from the analysis of the 11–14 μm AIB spectra in the Orion Bar.[18] The interstellar PAH family is dominated by compact PAH structures with long straight edges with few "corners" as in structure a. A minor contribution is due to PAHs characterized by more "bay" or "fjord" edges represented here by structures b and c. The compact PAHs are a factor ≃ 7 more abundant than the irregular PAHs of b and c. The number of solo (s), duo (d), trio (t), and quartet (q) groups are indicated in each panel. Figure taken from.[18] Reproduced with permission from Astronomy & Astrophysics, © ESO.

3 with the emphasis on validation of the calculated vibrational spectra of PAHs and on the effects of anharmonicity. Models for the IR cascade involved in the IR emission of isolated PAHs are described in Section 4. While it is generally thought that the interstellar PAH family is very diverse, the possible dominance by a few, very abundant, large PAHs in space—the so-called grandPAHs—is discussed in Section 5. Laboratory experiments have demonstrated the importance of electronic fluorescence for small PAH cations and this is discussed in Section 6. The topic of rotational transitions of PAHs is reviewed in Section 7 with an emphasis on the contribution of interstellar PAHs to the anomalous microwave emission (AME) and on the exciting identification of rotational transitions of PAH derivatives in the dark cloud core, TMC1. The diffuse interstellar bands (DIBs) are discussed in Section 8. Section 9 reviews recent progress in our understanding of the photochemistry and gas phase formation of PAHs under conditions relevant to space. This review is concluded with a summary and future directions (Section 10).

## 2. THE AROMATIC INFRARED BANDS

The early release science program, PDRs4All, led by Els Peeters, Emilie Habart, and Olivier Berne, has provided a wealth of data that illustrates well the power of the NIRSpec and MIRI integral Field units on JWST for the study of interstellar PAHs.[14] In this program, a strip of ≃ 3 × 25″ was mapped in spectroscopy (Figure 1), crossing the prototypical PhotoDissociation Region (PDR), the Orion Bar.[15] PDRs separate the ionized gas surrounding luminous hot stars from the surrounding molecular cloud. The Orion Bar is the bright PDR to the southeast of the Trapezium star cluster in the Orion Nebula and the edge-on geometry of the Orion Bar illustrates well the layered structure of PDRs, separating out the gas in the H II region—ionized by EUV photons ($h\nu > 13.6$ eV)—from the atomic zone, processed by far-UV photons $h\nu < 13.6$ eV.[15] These two zones are separated by a very thin ionization front (IF), where hydrogen goes from ionized to neutral atomic gas. Deeper in, hydrogen goes from atomic to molecular in the $H_2$ dissociation front (DF). Carbon is once ionized in the PDR but will recombine to C I and then to CO well beyond the DF.[15,16] PAHs in the PDR are electronically excited by far-UV photons and relax vibrationally by emission of one IR photon at a time, creating the AIBs. The results of the PDRs4All program have been published in some 15 papers (and counting). Here, I highlight a few of the important results.

The JWST results fully confirm and extend the incredibly rich spectrum of the AIBs (Figure 1). Besides the well-known IR emission features at 3.3, 6.2, 7.7, 8.6, and 11.3 μm, the observed interstellar spectra show a wealth of weaker features, including bands at 3.4, 3.5, 5.25, 5.65, 6.0, 6.9, 10.5, 11.0, 12.7, 13.5, 14.2, and 16.4 μm. Moreover, many of the well-known features show weak substructure (Figure 1).[13] For the Orion Bar, an extensive list of the features identified is available.[13]

While the JWST footprint is very small −3 × 3″ to 6.6 × 7.7″ with a pixel scale of 0.1−0.27″, depending on wavelength— adjacent "frames" can be very effectively stitched together and maps extending over ∼30″ scale-sizes can be created. This provides a very detailed view of the spatial variations of the AIBs. These variations can be linked to variations in the local physical conditions in the PDR through the rich set of atomic and molecular lines in these spectra.[12,17] Detailed analysis of the spatial behavior of the subtle spectral variations over a source allows for the decomposition of the AIBs in subcomponents and such decomposition studies are an active area of research.

As a first illustration of the power of the spectral surveys enabled by JWST, the analysis for the 11−14 μm range in the Orion Bar spectrum is discussed in more detail. The AIB spectrum in the 11−14 μm region is very rich with two strong features, at 11.2 and 12.7 μm and a number of weaker features and these split out into multiple subcomponents.[18] This spectral region is home to the CH out-of-plane bending modes of PAHs. Extensive laboratory and quantum chemistry calculations have revealed that the peak positions of these modes are very characteristic for the specific molecular edge structure of the emitting species[a] and the CH OOPS modes of isolated CH (solo), two adjacent CH (duos), three adjacent CH (trios), and four adjacent CH (quartets) on a ring separate out spectrally.[20,21] There is a slight dependence on charge state and on the size of the molecule. This spectral segregation has been extensively used in the analysis of astronomical spectra.[18,22] Analysis of the Orion Bar spectral imaging study[18] reveals that the main band at 11.2 μm shows two main components at 11.205—due to solo modes—while, following,[23] the weaker subband at 11.25 μm—mainly present deep in the Orion Bar PDR (i.e., DF3)—has been attributed to PAH clusters. The AIB at 12.7 μm separates out into six components mostly due to trio modes. The 12.0 μm AIB has six components and is mainly attributed to duo H in compact PAHs but there is a component associated with duo modes of PAH clusters also responsible for the 11.25 μm band. The contribution of duo modes to the 12.7 μm band on the other hand is minor.[18] The 13.5 μm band is due to quartets while the 14.2 μm band may be due to quintets (5 adjacent CH groups).[18]

Extending the earlier SWS/ISO analysis,[22] the JWST analysis demonstrates that the PAH family is dominated by compact PAHs with long straight edges and only a few "corners" (Figure 2).[18] There is some diversity in the types of corners as exemplified by the presence of duo, trio, and quartet modes in the spectrum and the presence of multiple components in these modes, but corners are few and compact, regular PAHs dominate the interstellar PAH family by a factor 7 over irregular PAHs. The composition of the PAH family varies somewhat with location in the Orion Bar PDR, but these variations are mainly limited to the "corner" structures; i.e., variations in the relative strength of the spectral components comprising the duo, trio, and quartet modes. Finally, as already indicated by the analysis of ISO spectra,[23] PAH clusters are present deep in the

Orion Bar PDR—c.f., the 11.25 μm band—but mostly limited to the H₂ dissociation front (DF3) and beyond.

The dominance of compact PAHs in the interstellar PAH family is supported by analysis of the 3.3 and 8.6 μm AIBs, and the IR activity in the 6−9 μm range. The local environment of the CH bonds can perturb the peak frequency of the stretching mode.[26−28] In particular, steric hindrance of H atoms across "bay regions[b]" widens IR activity to a large frequency range.[21,29] In contrast, for a compact species with long straight edges, there is little interaction between the stretching motions of CH groups on adjacent rings.[30,31] As an example, the symmetric, compact PAH, pyrene, exhibits IR activity over a much narrower frequency range than the noncompact PAH, chrysene (c.f., section 3.2; Figure 6). This is a very general result.[32] Likewise, the 8.6 μm AIB seems to be connected to the CH in-plane-bending mode in moderately sized, compact PAHs with long straight edges;[20,21,33] For small PAHs, this band is very weak; for large PAHs, bands in this wavelength range become very strong relative to the CC stretching modes at shorter wavelengths. Finally, the number of CC modes in the 6−9 μm range increases with PAH size. The number of IR active modes is, in addition, a function of the symmetry of the species. This is illustrated in Figure 3 with the measured spectra of three isomers of

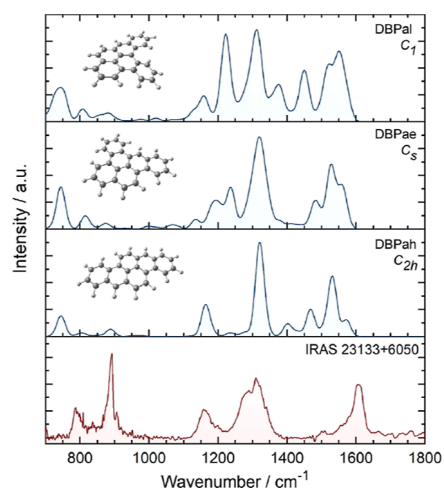

**Figure 3.** IR spectra of three, dibenzopyrene isomers ($C_{24}H_{14}$)[24] are compared to the AIB spectrum observed toward the HII region, IRAS 23133 + 6050.[25] The symmetry of these three isomers increases from top to bottom (dibenzo[a,l]pyrene (top) $C_1$, dibenzo[a,e]pyrene (middle) $C_s$, dibenzo[a,h]pyrene (bottom) $C_{2h}$). As this comparison illustrates, IR activity in the 6−9 μm (1100−1600 cm⁻¹) range strongly increases with decreasing symmetry of the species. The relative simplicity of the AIBs in this wavelength range points thus to a dominance of highly symmetric species in the interstellar PAH family. Figure taken from.[24]

dibenzopyrene of different symmetries[c], $C_1$, $C_s$, and $C_{2h}$: As the symmetry decreases, IR activity greatly increases.[24] While there are many weak AIBs in this wavelength range, the dominance of the 6.2 and 7.7 μm AIBs imply that the interstellar PAH family is dominated by moderately sized, highly symmetric, compact PAHs.[20,21,24,33,34]

As a second example of the power of JWST spatial-spectral studies, Figure 4 illustrates the results of a machine-learning analysis of the Orion Bar data,[35] revealing systematic variations in the relative strength of the 3.3 and 3.4 μm AIBs. The main band at 3.29 μm is due to the aromatic CH stretch. The satellite 3.4 μm AIB decomposes in multiple components, at 3.395, 3.403, 3.424, 3.48, and 3.52 μm, on top of a broad emission plateau.[12] The 3.4−3.5 μm components arise from CH stretching modes in −CH$_3$ and/or −CH$_2$ peripheral functional groups of PAHs or CH$_2$ groups created by superhydrogenation of the PAH C-skeleton, while the broad emission plateau has been ascribed to a blend of many CH/CC overtone and combination bands.[28,36−38] Ground-based observations established that the 3.4/3.3 μm AIB decreases systematically with distance in the PDR.[39,40] The JWST observations add to this by demonstrating that the subcomponents in the 3.4 μm AIB vary in tandem. Besides these obvious variation in the relative strength of the 3.4/3.3 μm AIBs, the 3.29 μm band shows subtle variations in the width which correlate with the trend observed in the relative strength of the 3.4/3.3 μm AIBs. Observationally, the fractional abundance of the aliphatic groups responsible for the 3.4 μm bands decreases relative to the aromatic CH group from the H$_2$ Dissociation Front (DF3) toward the IF.[35] This decrease with increasing intensity of the FUV radiation field has been attributed to the lower stability of a CH bond in aliphatic groups than in aromatic groups.[39,40] The photochemical evolution of PAHs is a balance between photochemical loss of side groups/substituents and chemical reactions of the resulting radicals with gas phase species and this balance tends to shift toward preferential loss of aliphatic groups in the increasing radiation field near the PDR surface.[40] As photochemical rates are a strong function of the internal excitation[41]—exponentially decreasing with size—the concerted behavior of the 3.395, 3.424, 3.48, and 3.52 μm (but not the 3.403 μm) subcomponents in the 3.4−3.5 μm implies that their carriers originate from the same type of side group attached to very similar-sized PAHs.[35]

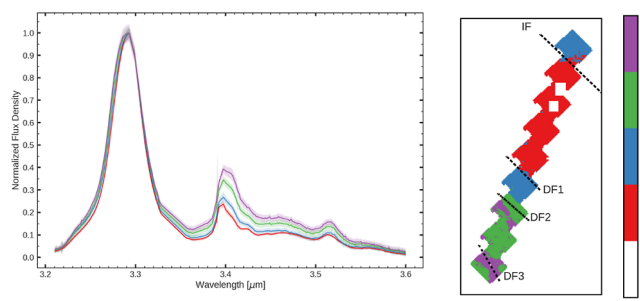

**Figure 4.** Machine learning techniques provide a powerful tool to extract systematic variations in AIB profiles as illustrated here for the CH stretching region in the Orion Bar data. Average spectral profiles (left) and spatial footprints (right) for all components in the 3.2–3.6 μm region have been extracted from the JWST spatial-spectral survey of the Orion Bar using a machine learning technique.[35] Colors are coordinated in the left and right panels. Shaded area around the spectral profiles indicate the 1 σ variations over the spatial region. Each spectrum is normalized to the measured peak intensity of the 3.3 μm AIB. However, the intensity of this band varies systematically by a factor 2 with depth in the PDR from the ionization front (IF) to the H$_2$ dissociation fronts (DF). Figure taken from.[35] Reproduced with permission from Astronomy & Astrophysics, © ESO.

As a final illustration of the power of spatial-spectral studies with JWST, detailed MIRI observations of the C-rich planetary nebula, NGC 7027, provide new insight in the structure of the C-skeleton of PAHs. ISO/SWS studies revealed detailed variations in the profiles of the AIBs. Based on these early studies an AIB classification scheme—class $\mathcal{A}$ to $\mathcal{D}$—was developed, linking profile and peak position of the individual AIBs.[25,30,44,45] Typically (but not always), classifications of the different AIBs are correlated and are linked to object type: Class $\mathcal{A}$ in the ISM and class $\mathcal{B}$ in (some) Herbig AeBe and T-Tauri stars, post-Asymptotic Giant Branch objects and Planetary Nebulae[d]. Some sources—notably NGC 7027—showed a mixed profile, suggesting an evolutionary link between the different classes, where, supposedly, class $\mathcal{B}$ carriers are injected into the ISM by C-rich AGB objects where they are processed over long times by the strong UV radiation field to class $\mathcal{A}$ carriers. JWST/MIRI observations have spatially resolved the transition from class $\mathcal{A}$ to $\mathcal{B}$ in NGC 7027 and these data allowed a decomposition of the 6.2 and 7.7 μm AIBs, each, into two components and the blue (6.205 and 7.6 μm) and red (6.26 and 7.8 μm) components correlate well but the blue components do not correlate with the red components.[43] The detailed IR activity, including the precise peak positions, in the 6−9 μm range depends strongly on the PAH size and molecular structure.[20,21,33] Figure 5 compares the characteristics for the blue and red components in the 6.2 and 7.7 μm AIBs in NGC 7027 with quantum chemical calculations for a select group of PAH cations.[42,43] As this figure illustrates, the details of the IR activity in the 6−9 μm range are unique to specific PAHs and provides a powerful tool to identify AIB carriers. At this point, it is unclear whether the red and blue components identified in NGC 7027 represent the two end members in the 6.2 and 7.7 μm profiles. Likewise, at this point, the IR activity of only a small subset of relevant PAHs has been investigated. Nevertheless, it is clear that these studies have great potential for the identification of the specific PAHs present in space.

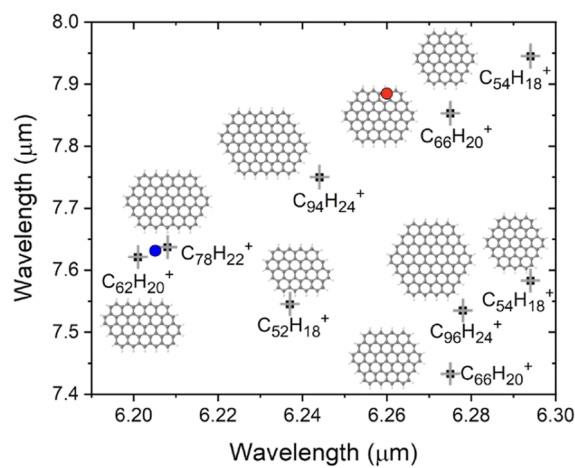

**Figure 5.** 6.2 and 7.7 μm peak positions of 7 cationic PAH species (data taken from.[42] Guided by existing analysis of astronomical AIBs, selected species have been limited to compact PAHs with a long, straight-edge molecular structure. Uncertainties for the theoretical band positions are shown by the gray error bars. Observed peak positions of the blue and red components of the 6.2 and 7.7 μm bands in NGC 7027 are indicated as blue and red circles. Figure taken from.[43] Reproduced with permission from Astronomy & Astrophysics, © ESO.

As these examples illustrate, the JWST IR spectrographs provide a powerful probe of the composition of the interstellar PAH family and observed spectral variations can be linked to the local physical conditions. This has the potential of providing an astronomer's tool to probe the astronomical conditions and evolutionary status of distant regions through observations of the PAHs and this is beginning to be explored.[46−48] Of course, this approach rests on a validation of the adopted relationship between spectral characteristics and the physical conditions established on nearby, well-resolved, and well-studied galactic regions and supported by laboratory studies and quantum chemistry calculations on PAH spectral characteristics and by astronomical model studies of the processes involved in their evolution in the ISM.

## 3. INFRARED SPECTROSCOPY OF PAHS

### 3.1. Harmonic Theory and Its Challenges

A variety of laboratory techniques have been used to measure the infrared spectra of PAHs (c.f., Tielens[1] and references therein). As experimental studies of individual species are very demanding, the field has focused on quantum chemistry studies using density functional theory in the harmonic approximation. The level of theory employed (i.e., basis set and functional) is balance between computational accuracy and speed. Despite advances in computer technology, the calculation of large PAHs remains challenging and only a very limited set of large, regular PAHs have been investigated. The results of a homogeneous set of calculations using B3LYP/4−31G have been made available to the community as a database, PAHdb, on a dedicated Web site (https://www.astrochemistry.org/pahdb/theoretical/3.20/default/view) maintained by Christiaan Boersma at NASA Ames Research Center.[19] A similar database (https://astrochemistry.oa-cagliari.inaf.it/database/) is maintained by G. Malloci and G. Mulas at the astronomical observatory of Cagliari. The PAHdb provides a visualization tool of the motions involved in the vibrational modes. Astronomical tools allow weighted coadding and stacking of the spectra in the PAHdb, fitting of astronomical spectra, and even the use of emission spectra at different temperatures. This database has been widely used to analyze astronomical vibrational spectra.[49−52]

These quantum chemical spectra provide an easy approach to analyze astronomical observations into a broader framework and address issues such as the ionization fraction, the size, and the molecular structure of the interstellar PAH family. However, the limitations of the quantum chemistry methods have to be kept in mind and this approach can only identify and classify trends in the data and do not necessarily provide accurate values for the parameters investigated. Three examples serve to illustrate this issue.

First, the quantum chemistry spectra have been calibrated against experimental spectra of a small set of PAHs with sizes $\lesssim$ 40 C atoms. These comparisons revealed a systematic mismatch in peak position due to anharmonicity and basis set errors but shifting by a factor of $\simeq 0.96$ brings good agreement between theoretical and experimental spectra.[53,54] In more recent studies the agreement between theoretical and experimental peak positions has been further improved through the use of multiple scaling factors.[19] In assessing this issue, it also has to be kept in mind that the experimental bands were typically measured in low temperature matrices and that technique tends to introduce peak shifts, broadening of the bands, and intensity changes. Anharmonicity effects and the accuracy of harmonic calculations will be further discussed in Section 3.2.

Second, early on, it was recognized that these quantum chemistry methods overestimate the intrinsic strength of the 3.3 $\mu$m CH stretching mode relative to the CH out-of-plane bending mode.[53] The observed ratio of the 3.3/11.2 $\mu$m band in interstellar spectra is a strong function of the excitation of the emitting species and this ratio has been widely used to determine the size (i.e., number of vibrational modes) of the emitting

PAHs.[36,46,55−57] Recent analysis implies that, indeed, calculated CH stretch/CH out-op-plane bending ratios in the PAHdb are overestimated, typically by 34%.[58] As a result, while the derived trends in PAH size variations implied by variations in the 3.3/11.2 AIB ratio can be trusted, the precise values of the inferred sizes are not accurate. As an example, the observed ratio, 1.75, of the 11.2/3.3 $\mu$m AIB in the PDR of the reflection nebula NGC 7023[55] implies emission from PAHs with $N_C \simeq 40$ C atoms rather than 50 C atoms.[58]

Third, the 6.2 $\mu$m AIB is generally attributed to the CC stretch in PAHs.[36] Most sources have observed peak positions in the range of 6.2−6.3 $\mu$m, although there are some sources with peak positions as low as 6.4 $\mu$m.[25] Early experimental and quantum chemistry studies indicated that PAHs can explain peak positions around 6.3 $\mu$m, but fail to account for positions as blue as 6.2 $\mu$m. Subsequent experimental and quantum chemistry studies on polycyclic aromatic nitrogen heterocycles (PANHs[e]) reveal that as the N atom shifts to the interior of the ring, the peak position shifts toward the observed peak position of the 6.2 $\mu$m AIB.[59] This has been taken to imply that N-substitution deep in the PAH C-skeleton is a key aspect of the interstellar PAH family.[49,59,60] However, a more recent quantum chemistry study, using basis sets designed to account for polarization effects, concludes that the peak position of the CC stretch in large, compact PAHs span the 6.2−6.4 $\mu$m range, and the peak position is sensitive to the molecular geometry (see Figure 5 and Section 2).[42,43] Moreover, this study concluded that PANHs will show a strong 11.0 $\mu$m CH out-of-plane bending mode. Analysis of observations imply then that PANHs can contribute at most $\simeq 10\%$ of the observed 6.2 $\mu$m band and hence are at best a minor component of the interstellar PAH family.[42]

As these three examples demonstrate, harmonic calculations have issues and care should be taken when comparing to observed interstellar spectra. Benchmarking and experimental validation of quantum chemical spectra are key to such endeavors.

### 3.2. Anharmonic Theory

Anharmonic interaction between the different vibrational modes is a key uncertainty affecting their peak position. Quantifying this interaction is therefore an important step in validating quantum chemical models for the vibrational spectra of PAHs. Due to interaction with spectator modes, anharmonicity will lead to a broadening and shift of the bands observed in low temperature absorption experiments or calculated by DFT theory. The transition energy, $E$, between adjacent levels in a given mode $k$ is given by

$$\frac{\Delta E_k(\{n\})}{\hbar} = \omega_k + 2x_{kk}(n_k) + \frac{1}{2}\sum_{i\neq k} x_{ik} + \sum_{i\neq k} x_{ik}n_i \quad (1)$$

with $\omega$ the harmonic frequency, $x$ the anharmonic constants, $n$ the number of quanta in the vibrational levels, $n_i$ represents the spectator modes, and $n_k$ is relative to the upper level. The magnitude of the shift will thus depend on the degree of population of the spectator modes and the anharmonic constants of the interactions. At low temperatures, the vibrational levels are not populated and the last term can be ignored.

Validation of the theoretical spectra of PAHs is of prime importance for the analysis of the AIBs. As mentioned above, the scaling factor(s) introduced to match theoretical, harmonic peak positions with experimental peak positions are based upon comparison to matrix isolation spectra. However, interaction with the matrix will introduce shifts in peak position, will affect band profiles, and will change relative intensities. Specifically, measured peak positions are only accurate to $\simeq 6$ cm$^{-1}$.[62] Proper validation requires low temperature, high resolution spectroscopy of gas phase PAHs and comparison to high level, anharmonic quantum chemistry calculations and such studies have been undertaken for a set of well-selected, small PAHs.[26−28,31,38,61,63,64] Illustrative results are shown in Figure 6.

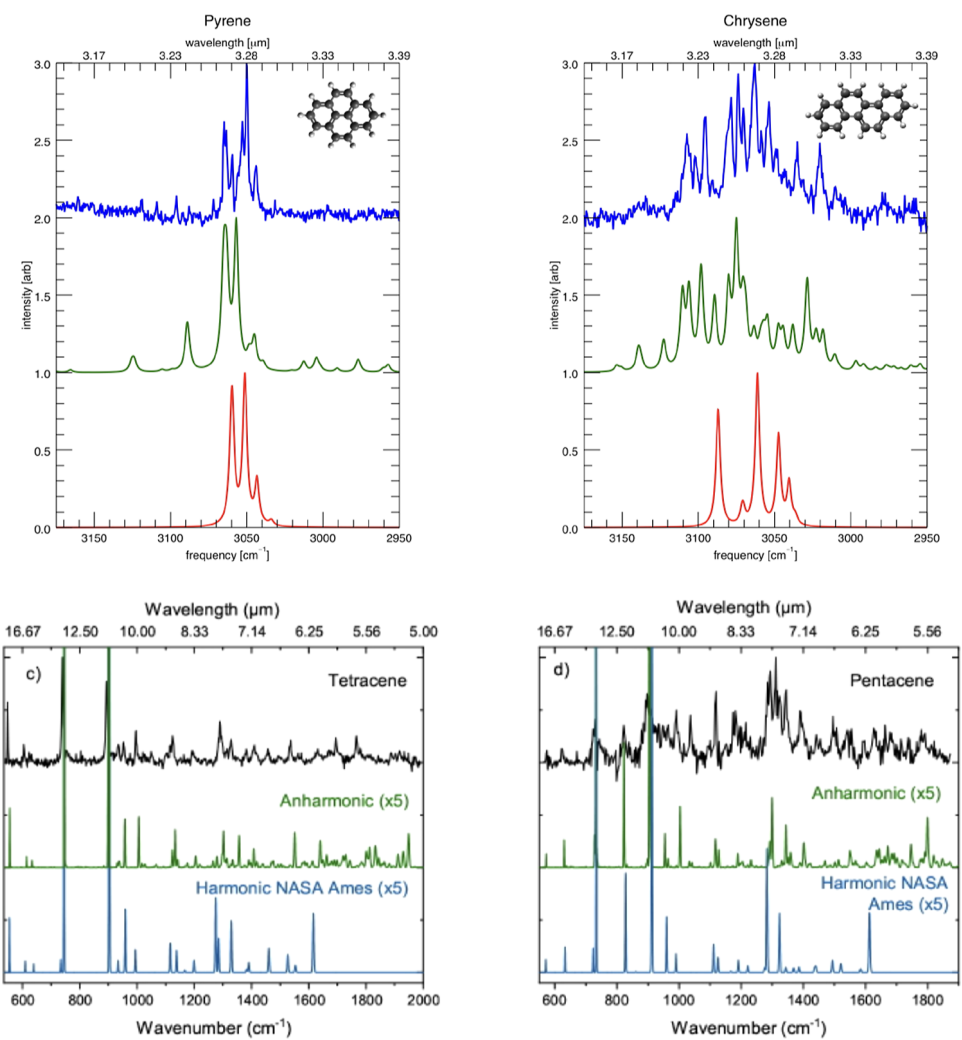

**Figure 6.** High resolution, low temperature absorption spectra of PAHs. Top panel: the CH stretching region of pyrene ($C_{16}H_{10}$, left) and chrysene ($C_{18}H_{12}$, right). Bottom panel: the CC stretching region and the overtone and combination region of tetracene ($C_{18}H_{12}$, left) and pentacene ($C_{22}H_{14}$, right). The three traces in each panel represent: Top trace: low temperature, gas-phase absorption spectra measured using a molecular beam technique. Middle trace: computed absorption spectra accounting for anharmonicity and resonance effects. Bottom trace: computed absorption spectra in the double harmonic approximation, with frequencies scaled by a factor 0.963. Top panels taken from.[32] Copyright [2022] American Chemical Society. Bottom panels taken from.[61] Reproduced with permission from Astronomy & Astrophysics, © ESO.

The experimental spectra use an ion-dip technique: IR-UV double resonance laser spectroscopy coupled with mass-resolved ion detection of mass- and conformation-selected PAHs seeded in molecular beams at a resolution of 0.1 cm$^{-1}$ in the CH stretching region ($\simeq 3000$ cm$^{-1}$) and $\simeq 5$ cm$^{-1}$ at lower frequencies (for details see[26]). The anharmonic theoretical spectra use the software package Gaussian16 to optimize the geometry of the molecules and to calculate the quartic force fields and IR intensities. The B3LYP hybrid functional and the polarized double-$\zeta$ basis set, N07D, are used as these are well suited for anharmonic calculations of medium-size molecules.[65]

The large number of mutually resonating modes are handled through a modified version of the SPECTRO software package to perform second order vibrational perturbation theory treatment (for details see[63]).

The results of this comparison show that the anharmonic peak positions in the CH stretching region agree with the high resolution experiments to 0.1%.[38,61,63,64] Experiments at longer wavelengths use the FELIX IR laser with a bandwidth of 1% of the frequency and agreement in peak position is at the 0.5% level (but that may be mostly linked to the uncertainty in the experimental peak positions). In contrast to the harmonic calculations, in neither wavelength range is a correction factor needed as the deviations have an approximately Gaussian distribution. We call out the good agreement in the 3000 cm$^{-1}$ region, both in terms of peak position and the range over which absorption appears. This is even more impressive as the CH stretching region is very susceptible to resonance effects. The anharmonic calculations also do well for combination bands and overtones in e.g., the 2000−2500 cm$^{-1}$ region. We note that harmonic calculations do reasonably well in peak position for the fundamental modes after a correction factor is applied—of 1.3% above 1000 cm⁻¹ and of 0.6% below—but they cannot account, of course, for combination and overtone bands.[61,64] Finally, the relative, integrated strength of the anharmonic calculations agree to better than 15% with the experimental values. As PAH emission spectra mainly depend on the relative strength of the modes and abundances are derived from UV absorption strength,[66] this is very satisfactory.

While the anharmonic DFT studies reach impressive agreement without scaling even in regions dominated by resonance effects, challenges still remain.

Specifically, the recent identification of PAH-nitriles through their pure rotational spectra[7,9,10,67] has awakened new interest in the infrared spectra of functionalized PAHs. In the context of this discussion, the calculated spectra of cyano-PAHs contain a clear warning signal. Experimentally, the C≡N stretching vibration falls between 2220–2240 cm⁻¹.[68] However, the large electronegativity of the CN group poses issues for quantum chemical calculations with commonly used basis sets and functionals. As an example, the C≡N stretching vibration occurs at 2229 cm⁻¹ for cyanobenzene, but calculations with B3LYP N07D including anharmonic corrections—which in general gives good agreement with experiments ($\pm$0.2%);[64] see Section 3.2)—predict a frequency of 2298 cm⁻¹; a difference of 69 cm⁻¹.[69] Likewise, experimentally, the CN stretch in 9-cyanoanthracene falls at 2207 cm⁻¹, while B3LYP N07D predicts this mode at 2282 cm⁻¹. Extensive analysis demonstrates that inclusion of higher electron correlation (using the basis set rDSD-TZ + B3LYP hybrid) is required to properly characterize the electronic structure of the highly electron withdrawing CN group on PAHs.[69]

The JWST spectrometers have a resolution of 1500−3500 (0.3−3 cm⁻¹), depending on wavelength and high S/N observations will allow peak position measurements to a fraction of this. On the theoretical side, more accurate peak positions will require improved basis sets and functionals. Likely, high order terms in the expansion of the potential energy surface are needed as well. Also, the measured absorption spectrum of e.g., naphthalene shows more absorption bands in the CH stretching region than theory can account for, implying that triple combination bands and overtones have to be accounted for as well.[63] On the experimental side, higher spectral resolution studies at longer wavelengths should be a priority to validate and guide the quantum chemistry results. Finally, because of experimental and computational limitations, only relatively small PAHs ($N_C \lesssim 30$ C atoms) have been investigated. As computer time increases with $N_C$ to the $4^{th}$ power, progress may require clever algorithms to extrapolate to PAHs with astrophysically relevant sizes ($30 \lesssim N_C \lesssim 60$ C atoms).

## 4. THE INFRARED CASCADE

### 4.1. Theory

In the ISM, the absorption time scale for UV photons by a PAH is $\tau_{uv} \simeq 1.4 \times 10^9/N_C G_0$ s, with $N_C$ the number of C atoms in the PAH and $G_0$ the strength of the UV radiation field in Habing units ($10^8$ photons cm⁻² s⁻¹;[70] or $\tau_{uv} \simeq 1000$ s in the Orion Bar. Collisional deexcitation is likewise a slow process, $\tau \simeq 10^9/n \simeq 2 \times 10^4$ s, at the low densities ($n \simeq 5 \times 10^4$ cm⁻³) of PDRs. In contrast, the IR vibrational cooling time scale is only $\simeq 0.1$ s. Hence, an interstellar PAH will be highly electronically excited after absorption of a UV photon. Internal conversion to the vibrational manifold of lower lying electronis states will occur on a time scale of $\sim 10^{-12}$ s and internal vibrational redistribution will "equilibrate" the energy over all modes on a time scale of $\sim 10^{-9}$ s. This population can be well described by a microcanonical temperature, $T_m$ (c.f., eq 4).[71] Finally, after absorption of a $\simeq 100,000$ cm⁻¹ photon, the highly excited PAH will rapidly cascade down in energy by emitting some 100 IR photons in vibrational transitions on a time scale of $\simeq 0.1$ s and it will stay "cold" until a new UV photon is absorbed after $\sim 10^3$ s.

The emission frequency of a given transition depends on the level populations of all modes (c.f., eq 1). Hence, accurate profile calculations require to keep track of the occupation numbers of all levels during the cascade. Given the large number of possible states in a given energy range, exact counting of the density of states is impossible. Therefore, Monte Carlo type samplings are used instead and the Wang−Landau method is the preferred method for calculating the density of states.[72] This has been shown to work well for PAHs[73] and this has been implemented in a cascade model.[32,74] An accurate profile calculation requires then the evaluation of $\sim 10^7$ UV absorption events and the emission of $\sim 10^9$ vibrational photons.

### 4.2. Validation against Absorption Spectra

With increasing internal energy the excitation of spectator modes[f] increases and, as a result, the anharmonic shift increases (c.f., eq 1). For low internal energies when only the lowest modes are populated, this will result in a spectrum with discernible multiple peaks separated by the anharmonicity constant (Figure 7). Each of the peaks in the theoretical spectrum (blue trace top panel Figure 7) represents discrete absorptions due to the fundamental with one or more excitation in one or more of the spectator modes. The assignments are based upon careful bookkeeping during the calculation. The excellent agreement between experiment and theory in the patterns of the absorption bands gives much confidence in the anharmonic calculations and the cascade model.

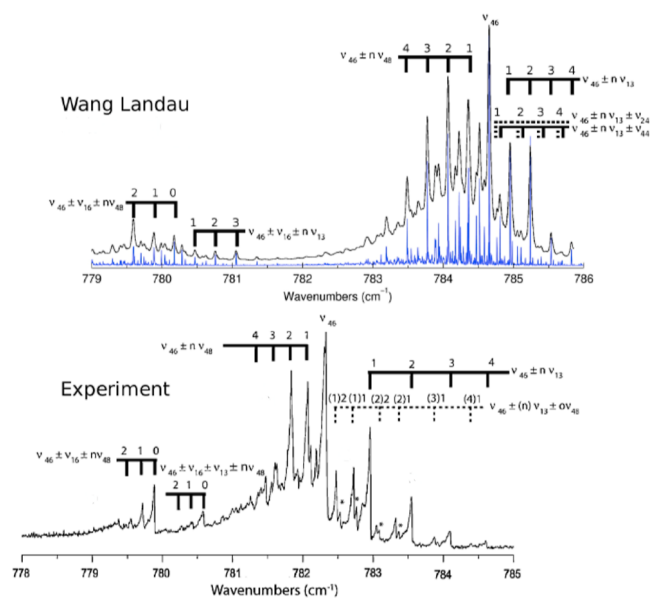

**Figure 7.** Comparison of the high resolution (0.005 cm⁻¹) experimental spectrum of the CH out-of-plane bending mode ($\nu_{46}$) of naphthalene measured at 300 K (internal energy $\simeq 0.04$ eV) (bottom trace;[75] and a model spectrum calculated, taking anharmonic interaction of this mode with other modes into account (top trace; blue with convolved spectrum in black;[74] Horizontal lines with tickmarks and labels indicate peak positions at different levels of excitation of relevant spectator modes. Adapted from.[74] Copyright [2018] American Chemical Society.

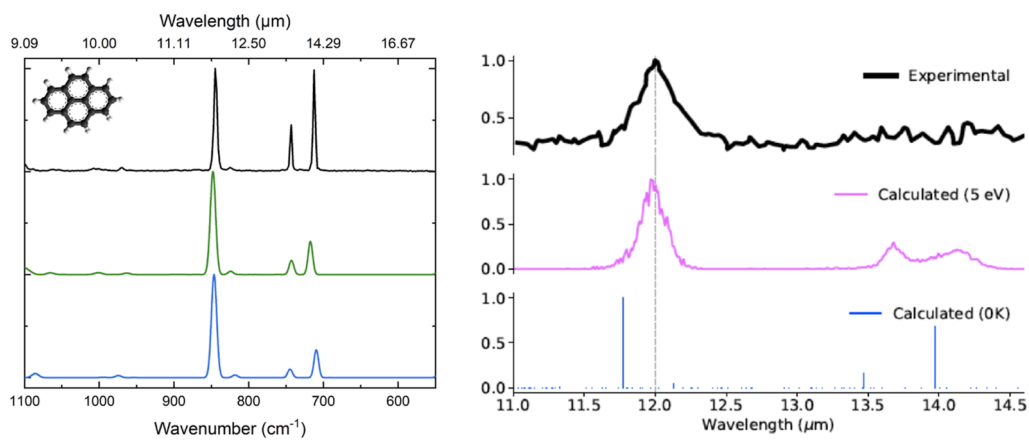

**Figure 8.** Left: comparison of measured and calculated low temperature absorption spectrum of pyrene ($C_{16}H_{10}$) in the region of the CH out-of-plane bending modes ($600-1100$ cm$^{-1}$). Top trace: Gas phase, ion-dip spectrum measured in a molecular beam.[76] Middle and bottom trace: calculated anharmonic and (shifted) harmonic absorption spectrum convolved with a 1% fwhm Gaussian, corresponding to the bandwidth of the FELIX laser.[64] Right: comparison of the measured (top trace;[77]) and simulated (middle trace;[78]) emission spectrum of pyrene with an internal energy of 5 eV. The bottom trace shows the low temperature, anharmonic absorption spectrum of pyrene as a stick diagram.[76] The shift and broadening of the emission spectrum at 5 eV relative to the calculated 0 K absorption spectrum reflects solely anharmonic effects caused by populated spectator modes; i.e., no additional broadening has been applied. Left panel reproduced by permission from.[76] Right panel taken from.[78] Reproduced with permission from Astronomy & Astrophysics, © ESO.

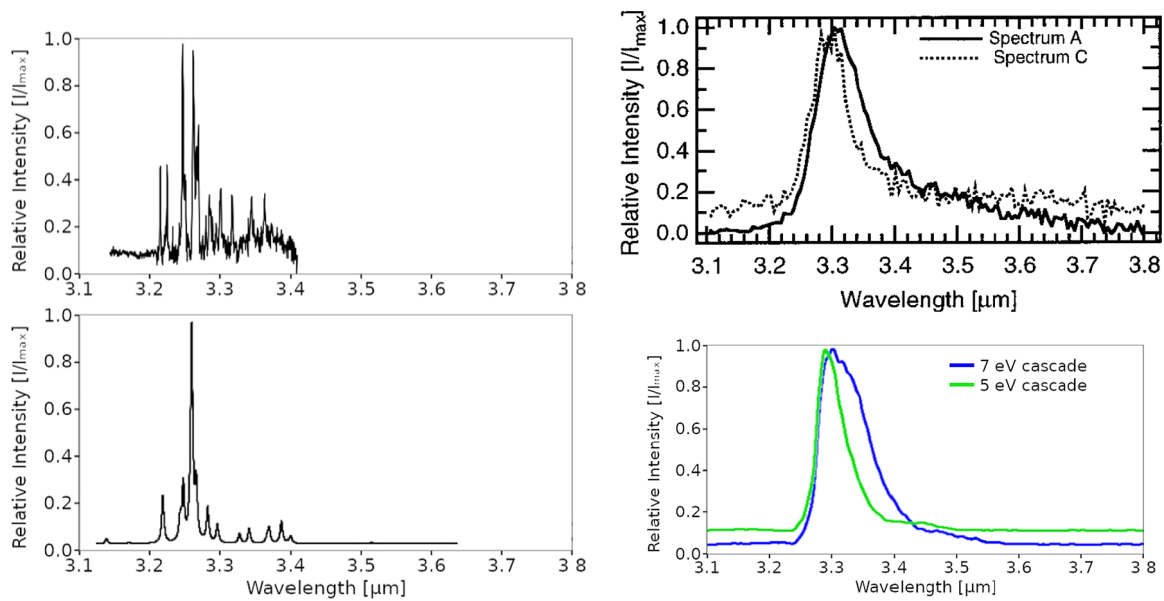

**Figure 9.** Left: comparison of measured and calculated low temperature absorption spectrum of naphthalene ($C_{10}H_8$) in the region of the CH stretching modes. Top panel: gas phase, ion-dip absorption spectrum measured in a molecular beam.[26] Bottom panel: calculated anharmonic absorption spectrum convolved with a 1 cm$^{-1}$ fwhm Gaussian.[63] Right: comparison of the measured emission spectrum of naphthalene at two different internal energies (5 and 7 eV, spectrum A and C respectively; Top panel[77]) and simulated emission spectrum of naphthalene at the same internal energies (5 and 7 eV) (Bottom panel;[32]). Left panels adapted from.[63] Copyright [2015] American Chemical Society. Right panels adapted from.[32] Copyright [2022] American Chemical Society.

### 4.3. Validation against High Temperature Emission Spectra

PAH emission spectra—highly excited by absorption of a UV photon—have been measured by the Saykally group.[77,79] At increased energies, more and more spectator modes will be involved, the peak position will shift, and the spectator modes will blend into a broad feature. Figure 8 compares measured and simulated spectra for the out-of-plane bending mode of pyrene.[61,77,78] The small shift of a few wavenumbers, between the calculated 0 K anharmonic absorption spectra and the measured, low temperature, ion-dip gas phase spectrum falls within the measurement uncertainties. With increased internal energy, emission peaks shift and the calculated anharmonicity effects agree well with the measured emission spectrum at the known 5 eV of internal energy. The simulated 12 $\mu$m emission band is somewhat narrower than the measured band. The experimental emission spectra also do not show the two bands at longer wavelength present in the measured and calculated, low temperature absorption spectra, perhaps reflecting the challenges in detecting such intrinsically weak bands. Overall, the agreement between measured and calculated anharmonic spectra is very satisfactory; the more as there are no free parameters involved. A somewhat different approach was used by[80] to simulate the thermal emission spectrum of pyrene at elevated temperatures. Equally good agreement is obtained with thermal experimental spectra over the full wavelength range.

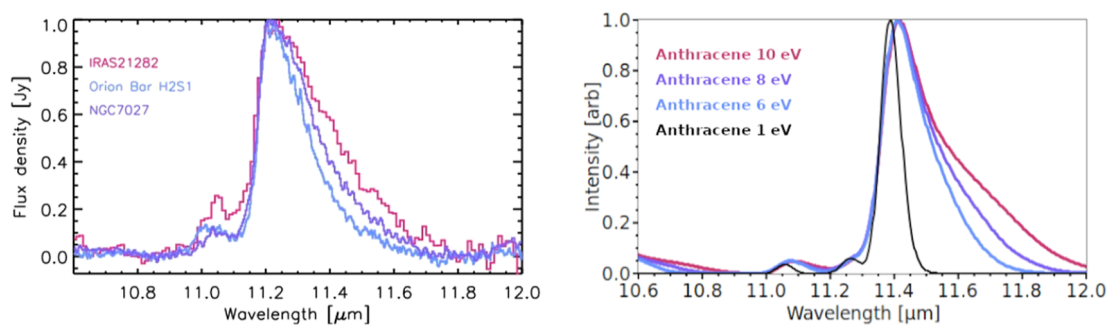

**Figure 10.** Left: observed profiles of the 11.2 $\mu$m AIB in the Orion Bar and the planetary nebulae, NGC 7027 and IRAS 21282 + 5050. Note how the strength of the red shaded wing of the 11.2 $\mu$m AIB increases between the sources. Figure adapted from.[25] Right: simulated cascade emission spectrum of anthracene at 1, 6, 8, and 10 eV. Note the change in the relative strength of the anharmonic, red-shaded wing as the internal energy increases. Adapted from.[32] Copyright [2022] American Chemical Society.

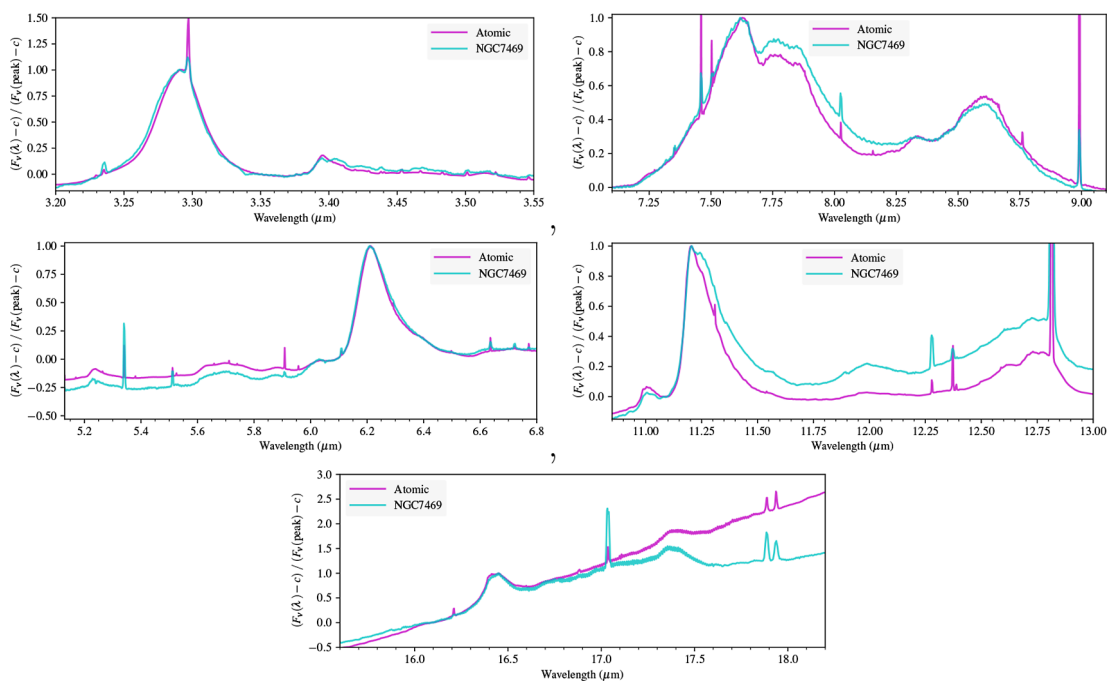

**Figure 11.** Comparison of the nearly identical profiles of the AIBs in the atomic zone (the brightest spot) of the Orion Bar and the starburst galaxy, NGC 7469. Figure kindly provided by Dr. Dries van de Putte.

Figure 9 compares measured and simulated spectra for the CH stretching mode of naphthalene.[26,32,63,77] The measured and calculated low temperature spectra show absorption due to fundamentals and combination and overtone bands. In the emission spectra at fixed internal energies of 5 and 7 eV, anharmonicity blends all of these bands into one feature and shifts it to lower frequency where the amount of the frequency shift depends on the internal energy. The blending into one feature, the anharmonic frequency shift, and the larger frequency shift for higher internal energy are well captured by the simulation. As Figures 8 and 9 demonstrate, the present theory describes the emission process and the resulting profiles at a level that is quite adequate for comparison to observations of the AIBs (see Figure 10).

## 4.4. The Anharmonic Profile

The observed profiles of the 11.2, 6.2, and 5.25 $\mu$m AIBs are characterized by a steep blue rise and a pronounced red wing (c.f., Figure 1).[81] While this may reflect blending of different bands into one AIB creating (fortuitously) a red wing, this type of profile is a telltale sign of anharmonicity.[32,82–85] The AIBs are the summation of the individual steps in the emission cascade from the initial high absorbed energy to low energy when the next photon is absorbed (Section 4.1). As the internal energy decreases, the population in the spectator mode decreases as well and the anharmonic shift will decrease, ending up, eventually at the low temperature absorption position. If the mode emits during the full cascade, this will lead to a red shaded wing. This is the case, for the strongest, low energy mode of neutral PAHs, the CH out-of-plane bending mode (c.f., Figure 10).[32,82] In contrast, for the strongest, high energy mode of neutral PAHs, the CH stretching mode, no red-shaded wing develops as the spectator modes are well populated whenever the CH stretching mode is populated enough to give appreciable emission.[32] The 3.3 $\mu$m AIB mode is therefore characterized by a redshift commensurate with its anharmonic constant, $\simeq 10$ cm$^{-1}$. Besides the 11.2 $\mu$m AIB, the 5.25 and 6.2 $\mu$m AIBs also show a pronounced red shaded wing. The anharmonic profile of these bands has not yet been explored in detail with a cascade model.

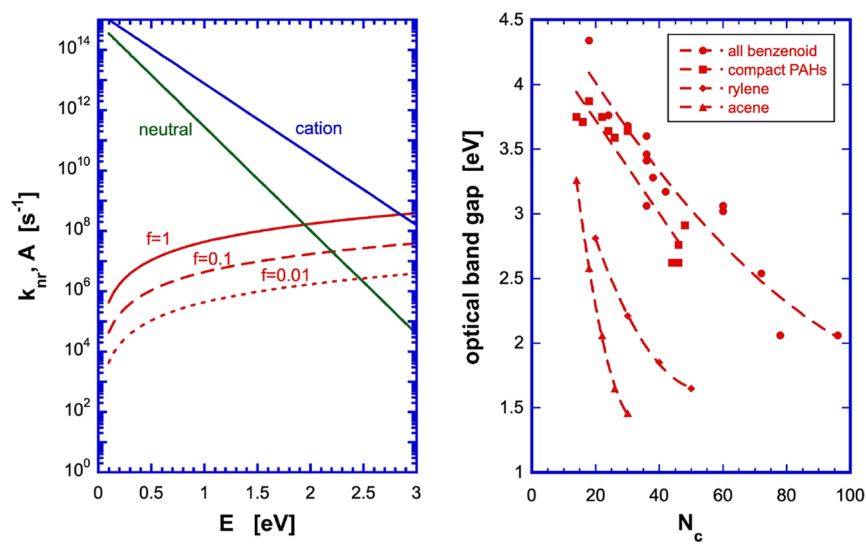

**Figure 12.** Left: curves representing nonradiative electronic relaxation rate, $k_{nr}$ of neutral (green) and ionized (blue) PAHs dominated by internal conversion (IC) are well represented by the energy gap law. Nonradiative rates based on the data and analysis in ref [89] The red curves are the radiative rates for three different oscillator strength spanning the relevant values for the $S_1 \rightarrow S_0$ transition in PAHs. Right: optical bandgap measured from the p-band transition in several series of PAHs,[h] illustrating the dependence on size (number of $\pi$ electrons) and molecular structure. Dashed curves represent least-squares fits to the different series. Adapted from the data presented in.[90]

## 5. GRANDPAH

Overall, the view of the field is that the interstellar PAH family is very diverse. However, there is evidence suggesting that the interstellar PAH family is dominated by a few, very stable PAHs—the so-called grandPAHs—that can survive the harsh conditions of the UV-rich environments of PDR surfaces.[86,87] First, this hypothesis is supported by the nearly identical spectra observed in very different galactic and extragalactic sources. This is illustrated in Figure 11 by the comparison of AIB spectrum of the brightest region in the Orion Bar PDR[13]—part of the well-know region of massive star formation in Orion at a distance of 410 pc—with that measured for the circumnuclear ring of star formation in the starburst galaxy, NGC 7469 at a distance of 50 Mpc.[88] While there are small variations in relative strength, the AIBs in these two sources are exceedingly similar in peak position and profile. Even the spectral substructure in the weakest AIBs repeats in detail[g]. This implies that the PAH family is exceedingly similar in these two environments. As a corollary, if the profiles of the AIBs were to represent blended components from many contributing species, we would expect to see spectral-spatial variations as the abundances of these species would be controlled by the local chemical history and local physical and chemical conditions.[86] Second, the CCC skeletal modes in the 14−18 $\mu$m region are very characteristic for the molecular structure. A large family of PAHs would give rise to a broad plateau but this spectral region has only two AIBs, at 16.4 and 17.4 $\mu$m (Figure 11).[13] Third, while there are quite a number of weak bands in the 6−8 $\mu$m region, this spectral region is dominated by 3 strong AIBs at 6.2, 7.6, and 7.8 $\mu$m. As this spectral regime is very sensitive to the molecular structure of the emitting species (Figure 5), this implies that a very limited number of species contribute. Finally, as discussed in Section 4.4, the red shaded profiles of the 5.25, 6.2, and 11.2 $\mu$m AIBs are a natural consequence of anharmonic interaction between different modes in a PAH, implying that these bands are not made up of a set of blended components from different PAHs.

In this view, the grandPAH hypothesis represents the "survival of the fittest" in the molecular Universe: an interstellar PAH family dominated by a few, extremely stable species controlled by photochemistry and kinetics. Spectroscopic analysis implies that the AIBs are carried by highly symmetric, compact PAHs and, within the class of PAHs, these are also the most stable species.[20,21,33] There is observational evidence for the weeding out of "weaker" members of the PAH family near the surfaces of PDRs in, for example, the loss of aliphatic groups and deuterated groups when PAHs approach the PDR surface. Likewise, the spatial variation in the 3.3/11.2 $\mu$m ratio reveals that the typical size of the emitting PAHs increases when approaching the illuminating star.[55] For NGC 7023, the abundance of PAHs decreases with decreasing distance to the star while the abundance of C$_{60}$ increases, indicative of a photochemical loss of small PAHs and a conversion of large PAHs into fullerenes.[6] In this respect, it should be recognized that, in the Orion Bar, as molecular cloud material flows through the PDR to the ionization front, a PAH typically experiences $\sim 3 \times 10^8$ photon absorptions of $\simeq 8$ eV, some $3 \times 10^4$ 2-photon absorption with a total energy of $\simeq 16$ eV, and $\simeq 4$ 3-photon events with a total energy of $\simeq 24$ eV. With these rates and internal energies, photochemistry of PAHs can clearly be expected to be important.

## 6. FLUORESCENCE

Electronic fluorescence is known to be an important process for small PAHs.[91,92] After the initial UV excitation to a highly excited electronic state—on a femtosecond time scale—internal conversion (IC) to lower lying electronic states will occur on time scales governed by Fermi's golden rule and characterized by an exponential bandgap law (Figure 12).[89,93−95] This is a convolution of two factors: the density of states increases exponentially with the energy, facilitating the transition. However, the Franck−Condon overlap integral decreases even faster with increasing energy. Best overlap is obtained for the highest frequency vibrations as fewer vibrational quanta are needed to match the energy of the excited electronic state.

At high energies, the density of electronic states is very high, nonradiative rates are very rapid, and IC dominates over radiative relaxation. However, for small PAHs, the energy gap between $S_1$ and $S_0$ is quite large and radiative relaxation becomes competitive (Figure 12). For some species, electronic fluorescence from $S_2$ to $S_0$ is competitive to IC as the larger bandgap is compensated for by a larger oscillator strength for the transition. The optical bandgap decreases systematically with PAH size (number of $\pi$-electrons)[h] and depends also on the molecular structure (Figure 12).[90] For ions, the radical nature implies that the energy gap between the ground and first excited doublet state is much smaller than for the corresponding neutral and the excitation will more quickly couple nonradiatively down to the ground state. Hence, for large neutral molecules, cations, and radical species, the band gap is typically small and the density of states is so high that transfer to the vibrational manifold of the ground electronic state is very rapid and relaxation through the vibrational modes will dominate over direct visible fluorescence.

Small neutral PAHs will also show phosphorescence. In phosphorescence, the electronic excitation transfers from the first excited singlet state ($S_1$) to the vibrational manifold of the lowest electronic triplet state ($T_1$), which lies below the first excited singlet state. In this transition, the electron spin flip is compensated by a change in the electron orbital momentum. The coupling factor of the spin-forbidden triplet–singlet transition is typically $\simeq 10^{-7}$ but this is easily compensated for by the much smaller energy gap between $T_1$ and that between $S_1$ and $S_0$. In a solid or liquid phase, collisions will then rapidly quench the molecule to the ground vibrational state of the triplet state from which phosphorescence to the singlet ground state ($S_0$) will occur. The time scale for phosphorescence is very long as a spin flip is involved ($\simeq 30$ s).[96] The singlet–triplet conversion is a reversible process and the forward and backward reaction are related through detailed balance, $k_b/k_f = \rho(E_S)/\rho(E_T)$. In the solid/liquid phase, transfer from $T_1$ back to $S_1$ involves vibrational excitation and is then sensitive to the temperature of the bath. This gives rise to Thermally Activated Delayed Fluorescence where the electronic energy stored in the triplet state is slowly released, controlled by the thermal energy of the system, and there is an extensive literature related to the development of efficient diodes.[97] As collisions play no role in the relaxation of PAHs in space, transfer to the triplet state will be followed by either the reverse reaction back to the vibrational manifold of the singlet state or by vibrational radiative relaxation in the triplet state. In the latter case, phosphorescence might occur. The vibrational spectra of PAHs in the triplet state have received scant attention in the astronomical literature and its importance for the AIBs is largely unexplored.[98] For large PAHs, transfer from $S_1$ to $S_0$ is very rapid and phosphorescence plays little role. For PAH cations, the first quartet state lies above the first excited doublet state and phosphorescence is also unimportant.

All internal conversion and intersystem crossing processes are reversible and much faster than radiative vibrational relaxation. Hence, the molecule will visit every electronic state many times and the probability to find the species in a given state is governed by their (relative) density of states. This can give rise to recurrent fluorescence from an excited electronic state to the ground electronic state.[101−104] The density of states in the excited electronic state will be much smaller than in the ground state as an appreciable amount of the excitation can be stored in electronic energy. However, this is compensated by the much higher Einstein $A$ value for electronic transitions (when Herzberg–Teller coupling is taken into account; see Section 8) than for vibrational transitions. As illustrated in Figure 13 for 1-cyanonaphthalene,[99] time scales for recurrent fluorescence can be much faster than vibrational radiative relaxation in the ground electronic state.[104]

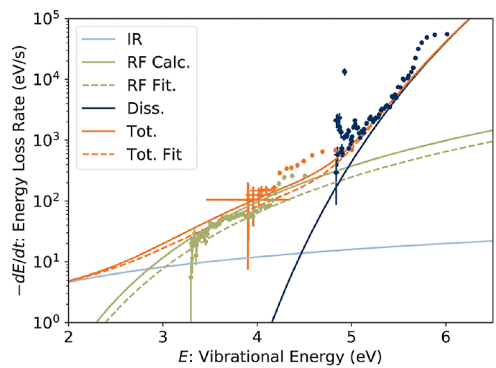

**Figure 13.** Energy loss rate for the 1-cyanonaphthalene cation as derived from the measured time evolution of the vibrational energy distribution of this species as it cools in the cryogenic storage ring, DESIREE.[99,100] Different colored symbols refer to analysis of different laboratory experiments. Solid lines are calculated energy loss rates due to dissociation and due to recurring fluorescence (RF) and vibrational (IR) radiative cooling. The dashed lines are a fit to the experimental data taking the oscillator strength $f$ of the RF transition as a free parameter. Reproduced from ref 99 with permission from the Royal Society of Chemistry.

The general importance of recurrent electronic fluorescence for the relaxation of highly excited small PAH cations has been demonstrated in a series of elegant experiments by Stockett and co-workers using the DESIREE storage ring facility in Stockholm.[99−103,105] As an example, for 1-cyanonaphthalene, recurrent electronic fluorescence dominates energy relaxation between $3 \lesssim E \lesssim 5$ eV. For higher energies, fragmentation dominates; for lower energies, IR fluorescence takes over. The importance of recurrent electronic fluorescence relative to vibrational emission is given by

$$\frac{I_{rf}}{I_{vib}} = \frac{A_{rf}}{A_{vib}} \frac{\rho(E - E_{rf})}{\rho_r(E - E_{vib})} \frac{E_{rf}}{E_{vib}} \tag{2}$$

with $E$ the total excitation energy, $E_i$ and $A_i$ the excitation energies and Einstein $A$'s relevant to electronic and vibrational excitation, and $\rho$ and $\rho_r$ the density of states of the parent and the reduced density of states of all vibrational modes except the one considered for IR emission. For large PAHs, the ratio of the densities of states can be approximated by[71]

$$\frac{\rho(E - E_{rf})}{\rho_r(E - E_{vib})} \simeq \exp[-(E_{rf} - E_{vib})/kT_e] \tag{3}$$

where $T_e$ is related to the microcanonical temperature

$$\frac{1}{kT_m} = \frac{d \ln[\rho(E)]}{dE} \tag{4}$$

through a correction for the finite heat bath of the species

$$T_e = T_m - \frac{E_{rf}}{C_m} \qquad (5)$$

with $C_m$ the microcanonical heat capacity; $C_m = dE/dT_m$. For this discussion, we can approximate this by[106]

$$T_e \simeq 11,000 \left(\frac{E(eV)}{3N-6}\right)^{0.8} \left(1 - 0.42\frac{E_{rf}}{E}\right) \qquad (6)$$

Hence, as the bandgap decreases, the probability to find the species in an excited electronic state will increase. The probability also increases when the internal energy increases. However, at high energies, ionization and fragmentation may take over as the dominant process (Figure 13;[100,107]). The bandgap energy depends on both size and molecular structure (Figure 12). The Einstein $A_{rf}$ for recurrent fluorescence will decrease rapidly (with $E_{rf}^3$). The Einstein $A$ also depends on the oscillator strength, which varies by 1−2 orders of magnitude between different species. Hence, further experimental studies are needed to assess the relative importance of these processes for astrophysically relevant PAHs.

Recurrent electronic fluorescence can have important implications for the analysis of the AIBs as well as for the assessment of the stability of PAHs in space.[100] Recurrent fluorescence will limit the typical excitation energy of small PAHs contributing to the AIBs and thus affect their infrared spectra; e.g., typical photon energies are $\simeq 6-8$ eV in PDRs, which is much larger than the energy where recurrent fluorescence starts to dominate the relaxation indicated by experiments on small PAHs (Figure 13). Specifically, this decrease in the typical excitation energy affects sizes derived from the observed ratios of the 3.3/11.2 $\mu$m AIBs and PAH abundances derived from observed ratio of the AIB intensity relative to the total infrared (dust) emission intensity.[66] Moreover, as the energy gap in cations is considerably smaller than for neutrals, recurrent fluorescence may affect the spectra of cations much more than those of neutrals. As a result, PAH ionization fractions inferred from observed 6.2/11.2 $\mu$m AIB ratios may have to be revisited. Furthermore, as the importance of recurrent fluorescence depends on the Einstein $A$'s of the electronic transitions involved and those can vary by orders of magnitude, the AIB spectrum may be skewed toward PAHs that are less susceptible to recurrent fluorescence. The more rapid electronic relaxation channel provided by recurrent fluorescence as compared to vibrational radiative relaxation will increase the stability of PAHs against UV-driven fragmentation. Recurrent fluorescence will also hamper generalization of experimental measurements of the fragmentation behavior of PAHs. Perhaps most importantly, direct measurement of recurrent fluorescence in space may provide a way to identify the presence of specific PAHs in space. For now, only the narrow emission features in the visible spectrum of the Red Rectangle point toward the importance of electronic fluorescence in space and these features may be related to the AIBs which are prominent in the 3−15 $\mu$m spectrum of that object. Further visible studies may be an interesting avenue to pursue.

# 7. ROTATIONAL TRANSITIONS

The rotational spectrum of PAHs has been discussed, in general terms[114] and for specific substituted PAHs (detected in TMC1).[7,9,10] Rotational excitation of PAHs in PDRs and in connection to the Anomalous Microwave Emission (AME) have also been discussed.[115,116]

Planar PAHs do not possess a permanent dipole moment and, hence, their rotational transitions are forbidden. Substitution, however, can break this symmetry and produce a nonzero dipole moment. The magnitude of the dipole moment will depend on the electronegativity/electropositivity of the substitution as illustrated in Table 1. For a highly electronegative group such as CN, the dipole moment of the substituted PAH can be substantial. The dipole moment will also depend on the size of the species.

**Table 1. Dipole Moments**[a]

| compound | substitution[b] | M [D] |
|---|---|---|
| pyrene | – | 0 |
| | $-CH_3$ | 0.51 |
| | $-OH$ | 1.62 |
| | $NH_2$ | 1.32 |
| | $-CN$ | 5.29 |
| | $-C_2H$ | 0.33 |
| | $-COOH$ | 2.36 |
| | $CH_2NH_2$ | 1.38 |
| | $-C_6H_5NH_2$ | 2.57 |
| | $-H^+$ | 1 |
| coronene | – | 0 |
| | $-CN$ | 5.67 |
| | $H^+$ | 3.5 |
| corannulene | – | 2.07 |
| sumanene | – | 1.96 |
| $C_{60}$ | – | 0 |
| | H | 0.3 |
| | $H^+$ | 3.6 |

[a]Data taken from[108−113]. [b]For pyrene, substitution at the 1 position. The change in symmetry—and hence the dipole moment—will depend on the substitution patters. Substitution at the 2 or 4 position will result in much larger dipole moments.

Substitution will lead to an asymmetric top molecule, the $J$ levels split, $K$ is no longer a good quantum number, and the energy level diagram and the selection rules are very complex. For example, for cyano coronene, the asymmetry parameter, $\kappa$, is $-0.12$,[10] which takes it very far from the prolate ($\kappa = +1$) or oblate ($\kappa = -1$) case. Because of the small rotational constant, $B \simeq 2 \times 10^{-3}(50/N_C)^2$ cm$^{-1}$, the partition function is very large even at low temperatures, $Z(T) \simeq 10^4(T)^{3/2}(N_C/50)^3$, resulting in low intensities of individual transitions and special filtering techniques are required to detect rotational transitions.[7,114]

The characteristic rotational temperature is, $\theta_{rot} = hcB/k \simeq 3 \times 10^{-3}(50/N_C)^2$ K and at 10 K, $J \simeq 50$ is energetically accessible. In a diffuse cloud, with $T \simeq 80$ K, levels around $J \simeq 160$ are within reach. For a PDR, the IR fluorescence process dominates, this simple assessment is less useful, and $J \simeq 250$.

The Einstein $A_{J,J-1}$ can be approximated by

$$A_{J,J-1} \simeq \frac{32\pi^4}{3hc^3}\nu^3\mu^2 \simeq 6 \times 10^{-9}\left(\frac{\nu_{rot}}{10GHz}\right)^3\left(\frac{\mu}{1D}\right)^2 \qquad (7)$$

where $\mu$ is the electric dipole moment and we have adopted $|\mu_{J,J-1}|^2 \simeq \mu^2/2(1-(K/J)^2) \simeq \mu^2/2$ appropriate for $J \gg 1$ and $J \gg K$. With a collisional deexcitation rate of $k_{dex} \simeq 3 \times 10^{-9}(N_C/50)^{1/3}(T/100K)^{1/6}$ cm$^3$ s$^{-1}$,[106] the critical density is $n_{cr} \simeq 100(\mu/\text{5D})^2$ cm$^{-3}$ for transitions around 10 GHz and relevant levels will be in LTE in molecular cloud cores. In diffuse clouds and PDRs, on the other hand, UV-pumped IR fluorescence will dominate the excitation process.

## 7.1. Anomalous Microwave Emission

Anomalous Microwave Emission dominates the emission spectrum of the Milky Way between 10 and 60 GHz. This emission is now generally ascribed to rotational transitions in large species. PAHs and related species may well dominate this emission. Consider a PAH in a PDR where excitation is dominated by UV absorption and relaxation is through ro-vibrational emission. The small difference in frequency, will favor $\Delta J = +1$ transitions but the slightly larger statistical weight will favor $\Delta J = -1$ transitions. The balance between these two will result in a Gaussian distribution in the rotational population[115,116]

$$\frac{n(J)}{n_0} \propto (2J+1)^2 \exp[-hcBJ(J+1)/kT_{rot}] \qquad (8)$$

which can be simplified to

$$\frac{n(J)}{n_0} \simeq \frac{4(J/J_{ir})^2}{\sqrt{\pi}J_{ir}} \exp[-(J/J_{ir})^2] \qquad (9)$$

where $T_{rot}$ is an effective rotational excitation temperature ($kT_{rot} = hcBJ_{ir}^2$) and $J_{ir}$ is the most probable rotational quantum number ($J_{ir} = (\bar{\nu}/6B)^{1/2}$) where $\bar{\nu}$ is the average energy of the emitted IR photons, $B$ is the rotational constant (typically, $B = 2 \times 10^{-3}(50/N_C)$ cm$^{-1}$), and the factor 6 accounts for the summation of the energies over the $K$ ladder. For a PAH, we can taken $\bar{\nu} \simeq 800$ cm$^{-1}$ and we have $J_{ir} \simeq 250$ and $T_{rot} \simeq 200$ K. For the rotational frequency, take $\nu_{rot} = 2BJ$ as a guide; i.e., $\nu_{rot} \simeq 35(50/N_C)(J/J_{ir}) \simeq 30$ GHz. In a PDR, $G_0/n_0 \simeq 1$, rotational deexcitation is driven by ro-vibrational emission and these estimates are reasonable.

Turning now to the AME: In diffuse clouds, collisions will become important, reducing the most probable rotational quantum number and, with some simplifying assumptions, this reduction can be assessed. For typical physical conditions in diffuse clouds, the most probable rotational quantum number ranges between 0.2 and 0.5 times $J_{ir}$ (Figure 14).[106] The frequency corresponding to the most probable $J$ is, $\nu_{rot} \simeq 15(50/N_C)$ GHz and the emission spectrum will peak at slightly higher frequency, $\simeq 25$ GHz. The intensity is given by

$$4\pi I_J \simeq 1.6 \times 10^{-23}\left(\frac{\nu_{rot}}{25\text{GHz}}\right)^4\left(\frac{50}{N_C}\right)^4\left(\frac{\mu}{1\text{D}}\right)^2 \text{ erg s}^{-1}\text{PAH}^{-1} \qquad (10)$$

Integrating this over the Gaussian rotational distribution results in

$$4\pi I_{rot} \simeq 6 \times 10^{-23}\left(\frac{50}{N_C}\right)^4\left(\frac{\mu}{1\text{D}}\right)^2 \text{ erg s}^{-1}\text{PAH}^{-1} \qquad (11)$$

At 22.8 GHz, the observed intensity of the AME is, $T_{AME}/I_d(100\mu m) = 10\ \mu$ K/(MJy/sr).[117] For the 100 $\mu$m dust intensity, we have from IRAS, $I_d(100\ \mu m)/N_H = 5 \times 10^{-21}$ MJy/sr/H-atom and the PAH abundance is $2.8 \times 10^{-8}(50/N_C)$,[106] resulting in $I_{AME} = 8 \times 10^{-23}(N_C/50)$ erg s$^{-1}$ PAH$^{-1}$. Hence, interstellar PAHs can account for much of the observed AME

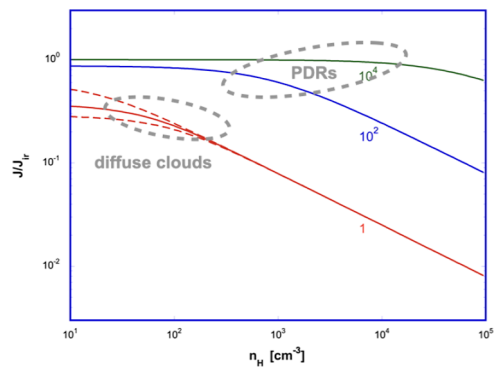

**Figure 14.** Rotational distribution of PAHs in the ISM can be described by a Gaussian distribution. The most probable rotational quantum number, $J$, relative to the most probable rotational quantum number when the IR cascade dominates, $J_{ir}$, is plotted as a function of density, $n_H$. Solid lines give the ratio for the values of the intensity of the UV radiation field in Habing units, $G_0$, and a dipole moment of $\mu = 1$ D. The dashed red lines illustrate the dependence on $\mu$ ($\mu = 0.1$ and 3 D). The approximate locations of PDRs and diffuse clouds are indicated by ellipses. Figure adapted from.[106]

emission if methyl groups or superhydrogenation induces a permanent dipole moment.

The dipole moment and IR emission properties of PAH clusters are very similar to those of PAHs and hence their rotational emission will also be similar. However, their abundance is lower by an order of magnitude[1] and they play less of a role. For very small grains with $N_C \simeq 10^3 - 10^4$ C-atoms, the abundance is lower by a factor of 100 compared to PAHs. Moreover, their peak emission will shift to lower frequencies, well outside the observed AME range and they are not relevant for the AME.

## 7.2. TMC1 Surveys

Over the last 5 years, deep surveys at centimeter wavelengths with the Greenbank (GBT) and Yebes telescopes[f] have revealed the presence of an unexpected reservoir of PAHs, their derivates, and related species in the cold dark cloud core, TMC1, through their pure rotational spectra. As PAHs have no permanent dipole moment, the emphasis has been on the substituted PAHs with e.g., cyanogen. Detected species include benzyne (o-$C_6H_4$), 1- and 2-cyanonaphthalene ($C_{10}H_7CN$), indene ($C_9H_8$), 2-cyanoindene ($C_9H_7CN$), 1- and 5-cyanoacenaphthylene ($C_{12}H_7CN$), 1-, 2-, and 4-cyanopyrene ($C_{16}H_9CN$), and cyanocoronene ($C_{24}H_{11}CN$).[7–10,67,118–121] Derived abundances are compared to those of related species in Figure 15. These abundances refer to the detected, and often substituted, species. Abundances of the PAH parent could be substantially larger. As the chemistry involving cyano substitution is likely very similar for all PAHs, the pattern may not change, though. We note that the abundance of species such as cyanopolyynes (HC$_n$N) or simple (hydro)carbon chains ($C_nH$) decreases in a systematic—sometimes alternating—manner with increasing number of C-atoms involved. In contrast, the abundance of small (cyano)-PAHs levels off at about $10^{-10}$. As small PAHs do not survive long in the ISM, the general consensus is that these PAHs (and related species) are formed inside dense clouds and that in this environment, small PAHs can survive long enough to build up an appreciable abundance.

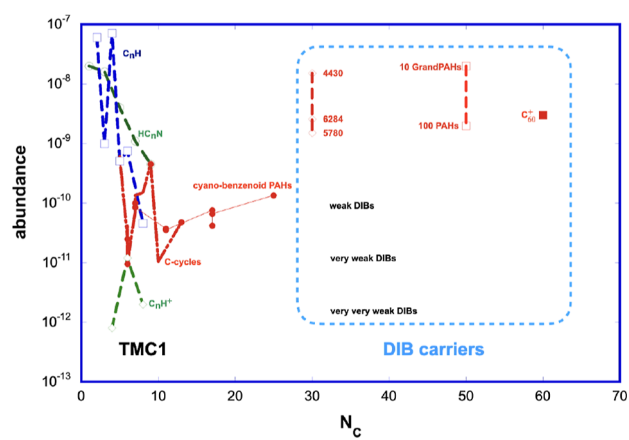

**Figure 15.** Left: abundances of PAHs and related species detected in the dark cloud core, TMC1, through their pure rotational transitions are compared to those of other hydrocarbons detected in this region. Note that the "parents" of PAH derivatives might have much higher abundances. Right: abundances of (potential) DIB carriers in the diffuse ISM. The abundance of $C_{60}^+$ is directly derived from its electronic transitions. Abundances for the carriers of the three strongest DIBs ($\lambda\lambda$ 4430, 6284, 5780 Å) have been calculated assuming an oscillator strength, $f$, of 0.1. The same $f$ has been adapted for the fiducial weak, very weak, and very very weak DIBs detected in the DIB spectra, but these transitions could also refer to intrinsically much weaker electronic transitions in more abundant species. GrandPAH abundances have been calculated assuming that 10% of the elemental C is locked up in 10 of these species. The 100 PAHs mark corresponds to the average abundance if the interstellar PAH family contains 100 species. Except for $C_{60}^+$, the placement of DIB carrier abundances along the $x$-axis are fiducial. Figure adapted from.[106]

## 8. THE DIFFUSE INTERSTELLAR BANDS

The diffuse bands, DIBs, are a set of several hundred absorption bands in the near-UV, visible, near-IR wavelength range ($\simeq 10,000-30,000$ cm$^{-1}$).[122,123] These bands are generally ascribed to electronic transitions in large molecules.[124] The field has been reviewed in two symposia.[125,126] In general, identification of the carriers of the DIBs has turned out to be very challenging as the list of potential candidates is large and laboratory studies are demanding. Elegant laboratory studies by Campbell et al.[5] have positively identified bands at 9365.2 and 9427.8 Å as electronic transitions in $C_{60}^+$. By now, a set of 5 bands in the far-red—at $\lambda\lambda$ 9365.2, 9427.8, 9577.0, 9632.1, and 9348.4 Å—are ascribed to this species.[127] The identification of $C_{60}^+$ lauds the dedicated efforts of John Maier and his group in Basel for over a decade to fine-tune the spectroscopic techniques to measure the absorption spectrum of this species. There are active experimental and theoretical programs to identify the molecular carriers of other DIBs, focusing on, among others, carbon chain radicals, cationic carbon rings, PAHs, and their derivatives, but none have hitherto been successful.[128–131]

The strength of DIBs relative to the local continuum is expressed as the equivalent width

$$W_\lambda = \int \frac{F_c(\lambda) - F_o(\lambda)}{F_c(\lambda)} d\lambda \tag{12}$$

with $F_c$ and $F_o$ the (adopted) continuum and observed flux, respectively. This can be written as

$$\frac{W_\lambda}{\lambda} = \frac{\pi e^2}{m_e c^2} f N \lambda \tag{13}$$

with $f$ the oscillator strength of the transition. DIB strengths are typically expressed relative to the column density of hydrogen nuclei, $N_H$, generally assessed from the measured color excess, $E_{B-V}$, or visual extinction, $A_V$, adopting $R = A_V/E_{B-V} = 3.1$, and $N_H/A_V = 1.9 \times 10^{21}$ cm$^{-2}$ magn$^{-1}$[32] where it should be understood that DIB strengths correlate reasonably well—but not perfect—with other tracers of the ISM. While there are some 500 DIBs catalogued, most of these are weak to very, very weak in strength ($W_\lambda/N_H = 10^{-23}-10^{-24}$ Å/H-atom). Indeed, there may be a DIB at every frequency at the very, very, very weak level.[133]

With such a high density of DIBs, wavelength coincidence does not suffice to positively identify a specific DIB with a molecular carrier. Support for an identification can come from the detection of multiple transitions due to, for example, vibronic progressions. Correlation studies have pursued this route to confirm or, more generally, refute molecular assignments. However, excitation of IR dark states populated by the IR cascade initiated by UV excitation may "spoil" perfect correlations between such sets of related bands[87] and this route may not be foolproof. For now, while there are many, many DIBs, identifications should perhaps focus on identifying the few strong ones; i.e., at 4430, 6284, and 5780 Å with strength of $W_\lambda/N_H \simeq 10^{-22}$ Å/H-atom. Adopting an oscillator strength of 0.1 for these transitions, abundances of $10^{-9}-10^{-8}$ are implied, suggesting that these molecular species are very abundant in the diffuse ISM—as abundant as typical hydrocarbon species in the shielded environment of dark cloud cores (Figure 15). Hence, DIB carriers must be very resilient molecules. As Figure 15 suggests, species such as the small PAHs or their derivatives detected in TMC1 might be plausible candidates if they could survive long enough in the diffuse ISM. Typical destruction time scales of PAHs in interstellar shocks are $\simeq 100$ Myr.[134] Hence, a rapid replenishment from dense molecular cloud cores is required. GrandPAHs could be considered as potential carriers of the strongest DIBs as well, as their inferred abundance is in the right range and they should be very resilient to destruction. Really small molecules, however, are unlikely candidates as they readily photodissociate in the interstellar radiation field and replenishment from molecular cloud cores would be too slow; e.g., abundances of $C_2H$ and c-$C_3H_2$ are down by one to 2 orders of magnitude in diffuse clouds as compared to the shielded environment of dense molecular cloud cores.[106]

At this point, it may be relevant to recall the visible spectrum of the Red Rectangle. This is a post Asymptotic Giant Branch (post-AGB) object where ejection during the AGB phase of stellar evolution has created a long-lived, dense disk of material surrounding the central binary object and an outflow directed into bipolar cones. The disk contains silicate dust while the outflow is dominated by PAH molecules.[135] The red color of this nebula is due to strong Extended Red Emission (ERE), an $\simeq 800$ Å wide emission band centered at $\simeq 6770$ Å.[136] This broad emission feature represents a luminescence process in nanograins and potential compounds include hydrogenated amorphous carbon and silicon. The ERE is a common characteristic of interstellar regions illuminated by far-UV radiation.[137] In the Red Rectangle, much narrower emission bands are superimposed on this broad luminescence band. These bands narrow and shift toward the red, toward the position of prominent DIBs.[138–142] However, the emission bands never reach the exact position of the corresponding DIB. As there are some 8 emission bands that link to DIBs, this cannot be a coincidence and the carriers of these emission bands and these DIBs must be the same. The variation in peak frequency and width of the emission bands in the Red Rectangle must then have a molecular physics origin. The frequency shift and profile narrowing of the emission bands might be related to a cooling of the rotational population of the carriers with distance from the star in the outflow.[139,140,143] When UV pumping dominates, the rotational population is set by the IR vibrational cascade after electronic excitation—leading to a Gaussian distribution centered at $J_{ir}$. This is (slightly) modified by collisional deexcitation of the rotational levels,[106,115,116] which will become relatively more important farther from the star (Figure 14), explaining the profile variations. The difference in the DIB (absorption) frequency and the emission frequency far from the star might then be attributed to a higher rotational temperature in the outflow than in the ISM, but that seems unlikely. Another possibility is a difference in the $^{12}C/^{13}C$ ratio between the absorbing species in the local ISM ($^{12}C/^{13}C = 70$)[144] and in the Red Rectangle (AGB and post-AGB $^{12}C/^{13}C$ ratios will be lower due to dredge up of nucleosynthesis products on the AGB).[145] For a $\simeq 30$–$50$ C-atom species, the species may shift from mostly pure $^{12}C_nH_m$ in the ISM to $^{13}C^{12}C_{n-1}H_m$ and the mass difference may induce a change in the difference between the vibrational frequency in the ground and the excited electronic state, resulting in a difference in the peak position. Alternatively, we note that many PAHs have low-lying electronic transitions that are Franck–Condon forbidden by symmetry but can become Herzberg–Teller allowed due to the dependence of the transition moment on the nuclear motions.[146,147] This results in a breaking of the mirror symmetry between absorption and emission.[f]

## 9. ORIGIN AND EVOLUTION OF INTERSTELLAR PAHS

PAHs are thought to be formed in the outflows from C-rich Asymptotic Giants Branch stars as molecular intermediaries in the carbon dust formation. While there is no direct evidence for PAHs in AGB outflows, likely, this reflects the absence of UV radiation that can pump the emission features. The C-rich AGB star, TU Tau, is an exception as a blue companion (spectral type A2) in this binary system provides sufficient UV to excite weak AIBs emission.[148] Indirect evidence for the formation of PAHs in C-rich AGB stars is provided by the analysis of graphite stardust grains isolated from meteorites whose isotopic composition betrays an origin in carbon-rich AGB stars. Laser-desorption laser-ionization mass spectrometry have revealed the presence of (some) small PAHs with an isotopic composition similar to that of the parent grain and hence formed in the same AGB outflow, stowed away and preserved in the parent grain during its sojourn in the ISM before entering the solar nebula and captured in the asteroid from which the meteorite was derived.[149] Finally, C-rich post-AGB objects and Planetary Nebulae (PNe)—the descendants of C-rich AGB stars—have UV-rich central stars illuminating the AGB ejecta and their IR spectra show bright AIB emission.[22,45]

PAH formation in the warm (1000–1500 K), dense ($\sim 10^{11}$ cm$^{-3}$) environment of AGB ejecta is akin to sooting chemistry of flames[150] and starts with the formation of benzene through the reaction of two propargyl radicals, $C_3H_3$ and through the reaction of buten-3-ynyl radicals, $C_4H_3$, with $C_2H_2$.[151–153] Subsequent growth to larger PAHs proceeds through repeated sequential steps of radical formation of the PAH through H abstraction followed by carbon (e.g., acetylene) addition, and, when appropriate, ring closure; the so-called HACA mechanism. Growth is then favored through the compact PAHs sequence of pyrene, coronene, ....[154] Rapid growth can then take place in the $900 < T < 1100$ K temperature range. For higher temperatures, both types of reactions (HA and CA) are reversible and entropy does then not favor growth. At 1100 K, the CA reaction becomes irreversible but HA is still reversible. This will drive rapid PAH formation and soot growth. Below 900 K, the HA reaction freezes out and PAH growth stops. At that point, PAHs will cluster and these clusters will then coagulate into carbon soot particles. Studies on the microscopic structure of soot particles produced in laboratory experiments show that the bulk of the molecular structures contain $\simeq 40$ C-atoms but species as large as $\simeq 250$ C-atoms are also present.[155] There is a second window of soot formation at higher temperatures ($T > 3500$ K), leading through fullerene structures, but that is of little relevance for circumstellar environments around AGB stars (or for sooting flames).

During the post-AGB and PNe phases of stellar evolution, PAHs formed during the AGB phase may be processed by the strong far-UV field of the central object as well as by shock waves driven by high velocity jets. This processing may, however, be limited to the innermost, circumstellar region and much of the $\simeq 1$ pc ejecta shell may be unaffected. In the ISM, PAHs will be processed by supernova driven shock waves at typical velocities of $\simeq 100$ km/s and by cosmic rays with typical energies of 10–100 MeV. Models predict a lifetime of PAHs in the ISM of $\simeq 100$ Myr.[134,156] As the PAH injection time scale by AGB stars is estimated to be 2 Gyr,[66] there has to be an efficient replenishment mechanism. This could be shattering of carbon soot particles by grain–grain collisions in shocks[157] or gas phase chemistry in dense cloud cores.[7]

### 9.1. UV Photochemistry

Photofragmentation is a major process in the UV-rich environments of PDRs and experimental studies have focused on elucidating the fragmentation pathways, their products, and the kinetic parameters involved.[158–167]

#### 9.1.1. Photoprocessing of Small PAHs.
For small PAHs, experiments reveal the presence of a number of competing pathways, including H, $H_2/2H$, CH, and $C_2H_2$ loss channels, resulting in the formation of smaller, dehydrogenated PAH species, C-rings, and C-chains. For small, cataconsensed PAHs, the H and $H_2/2H$ channels open up at slightly lower internal energies than $C_2H_2$ loss.[158,159] H-loss starts with roaming of an aromatic H, resulting in the formation of an sp$^3$ C with two H's.[168,169] H is then lost from these sp$^3$ sites. The experiments reveal a distinctive even/odd asymmetry as removal of the second H is facilitated energetically by the formation of a triple bond, resulting in a substantially lower energy cost.[161,170] The 2 H's can be lost either sequentially or as $H_2$ depending on the molecular structure of the species.[170] Barriers for H-roaming are $\simeq 3.2$ eV, while loss of $H/H_2$ from an sp$^3$ site has a barrier of $\simeq 2.5$ eV. As this is a two step process, the overall barrier for H/$H_2$ loss is 4–5 eV. These energies apply quite generally for small and large PAHs alike, except for PAHs with bay regions as the steric hindrance of the H's across the bay and the nonplanar structure of such PAHs results in energies lower by $\simeq 1.5$ eV.[171] The experiments confirm the expected, substantial kinetic shifts with PAH size, reflecting the increase in the degrees of freedom and hence the heat capacity.[158–160] As a result, in the ISM, H-loss from small PAHs can be single photon events but, for PAHs larger than $\simeq 30$ C-atoms, H-loss is driven by multiphoton processes.[41,170,172]

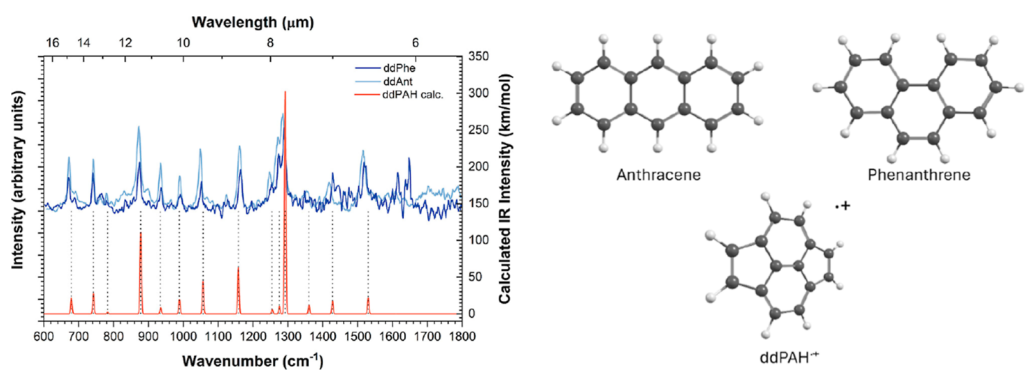

**Figure 16.** Left: IRMPD spectra of neon-tagged products of doubly dehydrogenated anthracene (light-blue) and phenanthrene (dark-blue) produced by 30 eV electron ionization. Bands beyond $\simeq 1660$ cm$^{-1}$ are likely artifacts. For comparison the DFT spectrum of pyracylene (red) is also shown, shifted by a factor 0.983 (to account for anharmonicity) and convolved with a Gaussian of width of 5 cm$^{-1}$ to mimic the experimental resolution. Right: structures of the isomeric $C_{14}H_{10}$ parents and their doubly dehydrogenated product, pyracyclene. Figure taken from.[174] Copyright [2025] American Chemical Society.

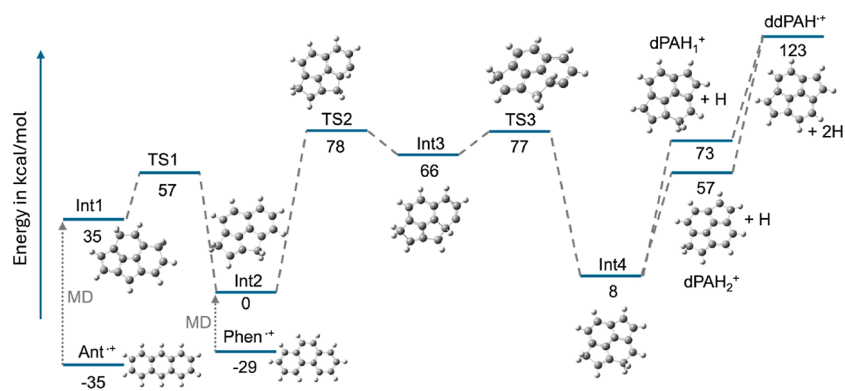

**Figure 17.** $C_{14}H_{10}^+$ potential energy surface identifying a reaction mechanism linking the parents with the daughter product, pyracyclene. The potential energy surface starts with intermediate structures identified as intermediaries in molecular dynamics simulations (Int1 and Int2)[175] and leads to the product identified using IRMPD spectroscopy (Figure 16). The transition state connecting Int1 and Int2 displays only the rate-limiting step (TS1). Figure taken from.[174] Copyright [2025] American Chemical Society.

For catacondensed PAHs, $C_2H_2$ loss has a slightly higher appearance energy[^f] than the H-loss channels by 0.3–0.8 eV.[159] This is different for pericondensed, small PAHs where much higher energies are involved (e.g., pyrene, 6 eV; coronene 9.1 eV).[167,173] This increase in energy reflects the difficulty in forming a four-member ring as, in contrast to small catacondensed PAHs, the compact structure of these species hinders adoption of the most favorable set of bond angles. However, even for these compact PAHs, the $C_2H_2$-loss channel will start to compete with H-loss channels at high enough internal energies. Also, after substantial dehydrogenation, the energies involved in $C_2H_2$ loss lower substantially, reflecting opening up of hexagons into large carbon-rings. For pyrene, the $C_2H_2$ barrier is only 3.94 eV after loss of 5 H's[173]. The large carbon rings formed in such highly dehydrogenated PAHs are susceptible to breaking off and fragments larger than 10 C-atoms will form C-rings, while smaller ones will form chains.[167]

Isomerization can compete very efficiently with fragmentation and this can have profound influence on the molecular structure of the interstellar PAH family. Figure 16 shows the infrared multiphoton dissociation (IRMPD) spectra of the doubly dehydrogenated products formed after electron ionization at 30 eV of the two $C_{14}H_{10}$ isomers, anthracene and phenanthrene. The IR spectra of the daughter products—trapped in the cryogenically cooled, 22-pole ion-trap, FELion and tagged with neon—have been obtained using the free electron laser for infrared experiments, FELIX, facility.[174] The measured spectra are essentially identical, revealing the formation of a common daughter product. Note also that the spectra are rather simple, indicating that this common species is highly symmetric. The measured spectra are compared to the harmonic DFT spectra calculated using the B3LYP/6−311++G(2d,p) functional and basis set and scaled to account for anharmonicity. This comparison reveals an excellent match between measured and calculated spectra. In contrast, calculated spectra of none of the direct dehydrogenation products of either isomer provide a match to the observed spectra. In addition, pyracylene is the lowest energy $C_{14}H_8$ isomer by 1.8 eV. These results demonstrate that these two PAHs isomerize after ionization and excitation to a common product and that this species is the highly symmetric, compact PAH, peracyclene, characterized by two pentagons.

Figure 17 illustrates a possible formation mechanism of the identified isomers.[174] Molecular dynamics studies revealed very similar intermediaries resulting from excitation of anthracene and phenanthrene, characterized by two adjacent pentagons and hexagons.[175] These two intermediaries are connected through a transition state (TS1) with an energy of ≃2.5 eV. Subsequent H migration to the tertiary C opens up the CC bond separating the 5−6 rings, which then closes to form a 6,5 rings structure (INT4). This results in an apparent migration/relocation of the 5 ring to the opposite side and a highly symmetric molecular structure. Only after this migration are the two H's lost from the structure.[174] The total energy involved is only 5.3 eV. Such isomerization pathways may be very common in the breakdown diagrams of PAHs, opening up new dissociation channels and resulting in more highly stable products that are not in the preferred bottom-up formation pathway. Note also that, in this particular example, the resulting species obeys the isolated pentagon rule—one reason for its stability—and may therefore be a potential precursor species in the bottom up pathway to $C_{60}$ formation.[174]

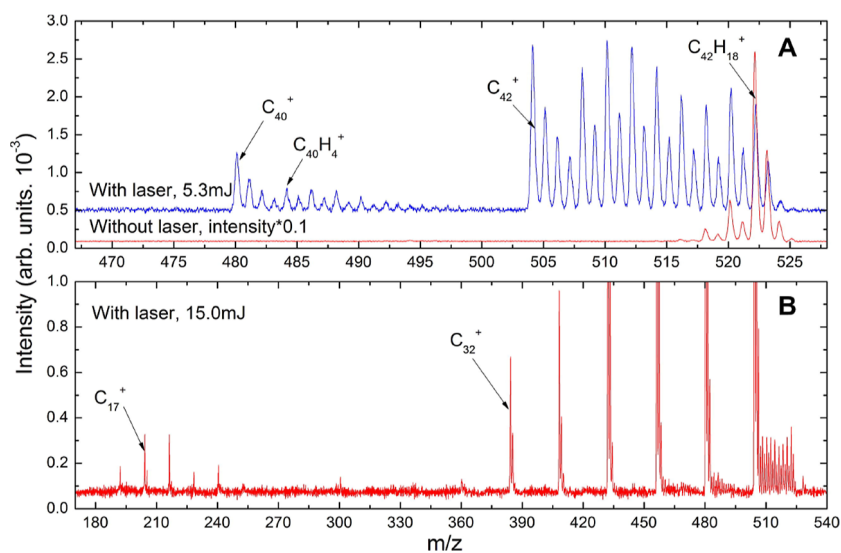

**Figure 18.** Photofragmentation of hexa-peri-hexabenzocoronene (HBC) cations trapped after evaporation and electron impact ionization. Panel A: time-of-flight mass spectrum of HBC radical cations without laser (scaled by a factor 0.1) and after irradiation at 532 nm with 5.3 mJ laser pulse energy. The predominant photolysis pathway involves sequential H-loss. Panel B: time-of-flight mass spectrum after irradiation with 15.0 mJ laser pulse energy. Once most of the H's have been removed through photolysis, the carbon structures evolve by losing $C_2$ units down to $C_{32}$. Smaller products involving rings and chains become accessible as well. Figure taken from ref 161.

Isomerization is likely a very general process in the photoprocessing of PAHs. In a similar study, after acetylene loss, the naphthalene cation undergoes facile 6- to 5-membered ring conversion to pentalene ($C_8H_6$).[176] Similar isomerization processes play a role in the photoprocessing of the $N$-containing PAH isomeres, acridine and phenanthridine ($C_{13}H_9N$).[177] In Section 9.1.2, evidence is reviewed for the isomerization of large PAH into cages and fullerenes after loss of all their H's.

**9.1.2. Photoprocessing of Large PAHs.** For large PAHs, photoprocessing starts predominantly with stepwise loss of H before C-loss channels become important.[161,162,178] Figure 18 illustrates this for hexa-*peri*-hexabenzocoronene (HBC; $C_{42}H_{18}$) cations.[161] As for small PAHs, H-loss has the characteristic even−odd pattern associated with the increased stability afforded by triple bond formation. For large PAHs, pure C-clusters are formed before C-loss starts. While these C-clusters might be small graphene flakes, it is quite likely that isomerization results in 3D C cage structures. Compared to small PAHs, the predominance of processing through the lowest energy channel (H-loss) for large PAHs reflects the increase in heat capacity with size as specific loss channels will open up at similar excitation temperatures for different PAHs. The microcanonical excitation temperature, $T_e$ scales approximately with $E/(3N-6)$ (c.f., eq 6)[106] and, as the size increases, more and more laser photons will need to be absorbed to reach the same temperature.

The photolysis of $C_{66}H_{26}^+$ follows the same overall pattern of almost complete H-loss before $C_2$ loss sets in and pure C-

clusters evolve. Figure 19 compares the fragmentation pattern of these C-clusters to those of the fullerenes, $C_{60}^+$ and $C_{70}^+$.[162]

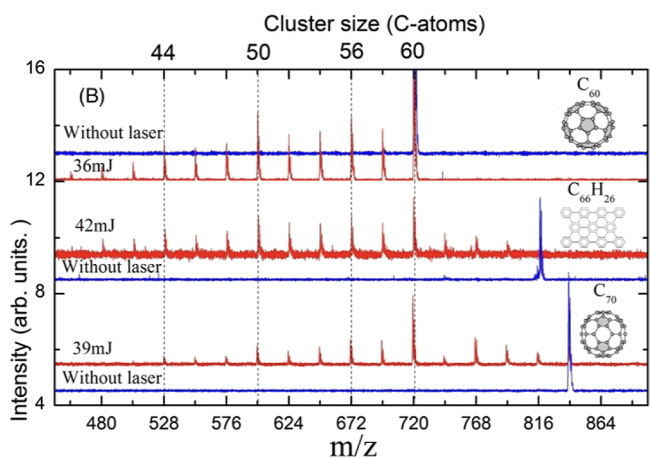

**Figure 19.** Comparison of the fragmentation pattern of the fully benzenoid cation, $C_{66}H_{26}^+$ and the fullerene cations, $C_{60}^+$ and $C_{70}^+$, irradiated at 355 nm. Note that the fragmentation pattern of the PAHs resembles that of the fullerenes. The "magic numbers" are marked by the vertical dashed lines and correspond to peaks with an $m/z$ of 720 for $C_{60}^+$, 672 for $C_{56}^+$, 600 for $C_{50}^+$ and 528 for $C_{44}^+$. Figure adapted from.[161] Reproduced by permission of the AAS.

Noteworthy, the pure C-cluster derived from this PAH reveals the same "magic numbers" as for the two fullerenes.[171]. Moreover, the abundance of the $C_{60}^+$ cluster in the $C_{66}H_{26}$ experiment can be greatly enhanced in photolysis experiments performed at 532 nm where $C_{60}^+$ does not absorb; a telltale signature of the formation of the $C_{60}^+$ fullerene. These experiments imply a 60% efficiency of $C_{60}^+$ fullerene formation.

**9.1.3. Astrophysical Implications.** In a PDR surface, the process of H-loss by photolysis will be balanced by reactions of the resulting radical with atomic H. In the UV-rich, surface layers of PDRs where $G_0/n_H \simeq 1$, UV photolysis will win for PAHs smaller than $N_{crit} \simeq 60$–$70$ C-atoms and they will be readily stripped of H.[41] The resulting highly dehydrogenated C skeletons are then further broken down through loss of $C_2$ units (Figure 20).[167,173] However, as the UV field is reduced by a factor $\simeq 30$ at the $H_2$ dissociation front deeper in the PDR ($A_V \simeq 2$ magn.), the balance between photolysis and H-reactions will "protect" even small PAHs against C-loss.[41] Thus, small PAHs are expected to be only lost from the interstellar PAH family—if initially present at all—near the PDR surface. Functional groups such as methyl ($CH_3$), hydroxyl (OH), methoxy ($OCH_3$), and cyanogen (CN) will effectively not be balanced by gas phase reactions as the gas phase abundance of these species is small in a PDR and, as their binding energy is less than that of aromatic H, their abundance will gradually decrease in the PDR as material is advected into the surrounding ISM through the PDR surface.[40] When the advected PDR material enters an HII region, photons as energetic as 30 eV become available, and C-loss channels for large PAHs become available as well and PAHs will be rapidly destroyed.

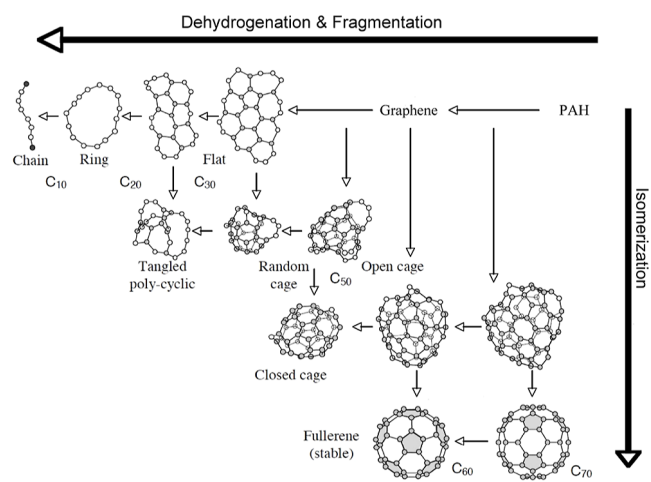

**Figure 20.** Schematic representation of top-down chemical evolution of PAHs in the ISM under the influence of UV photons, combining the effects of dehydrogenation and fragmentation with those of isomerization. Fully hydrogenated PAHs are at the top right side. Near bright stars, UV photolysis will preferentially lead to complete H loss and the formation of graphene-like structures. Further fragmentation may lead to the formation of rings, and chains. However, this process competes with isomerization to various types of stable intermediaries such as cages and fullerenes. Figure taken from.[6]

It has been suggested that UV photolysis of PAHs is the origin of small hydrocarbon radicals, such as $C_2H$ and $C_3H_2$, in PDR surfaces.[179,180] The high spatial resolution of JWST and ALMA allow a detailed comparison of the morphologies of the PAH emission and the hydrocarbon abundances. A detailed study of the Orion Bar[181] reveals that the $C_2H$ abundance peaks in the $H_2$ dissociation front and the measured abundances are well explained by pure gas phase models driven by the warm kinetic temperatures of the gas at these locations ($\simeq 500$ K) and the high vibrational excitation of UV-pumped $H_2$; both promoting rapid hydrocarbon formation from the prevalent $C^+$. The presence of small hydrocarbon radicals in the atomic zone closer to the PDR surface, on the other hand, may reflect the importance of UV-driven top-down PAH chemistry.[181]

## 9.2. PAH Formation in Dense Molecular Cloud Cores

As discussed in section 7.2, recent deep surveys at millimeter wavelengths have revealed the presence of small aromatic species in dense molecular cloud cores.[7−10,67,118−121] The focus has been on functionalized (e.g., CN) PAH derivatives as these have high dipole moments, but their presence implies a high abundance of the parent species. Taking benzonitrile as an example, formation is through reaction of benzene with the cyanogen radical. This is a rapid reaction at low temperatures with no discernible temperature dependence ($k \simeq 4 \times 10^{-10}$ cm³ s⁻¹),[183,184] implying a barrierless process. Likely, destruction of benzonitrile is dominated by proton transfer followed by dissociative electron recombination (the proton affinity of benzonitrile is 8.4 eV) and by absorption of UV photons—created by cosmic ray excitation of $H_2$—followed by fragmentation. In both cases, H-roaming is likely the first step followed by HCN elimination, resulting in benzyne ($C_6H_4$). With a CN abundance of $8 \times 10^{-9}$ in TMC1, the abundance of benzene is expected to be $\simeq 10^2$ times larger than that of $C_6H_5CN$ (i.e., $X(C_6H_6) \simeq 10^{-8}$). Likely, similar reactions—and enhancement factors—are involved for pyrene and coronene and their abundances are expected to be at levels comparable to $HCO^+$, HCN, $H_2CO$, and $CH_3OH$ in TMC1.

The presence of PAHs in dark cloud cores has renewed an interest in chemical pathways under ISM conditions and this has led to a multitude of experimental and theoretical studies.[185] PAH formation has to start with the formation of the first ring. Propargyl has been detected in TMC1 (abundance $9 \times 10^{-9}$)[186] and, in general, this radical is a key intermediary in the formation of PAHs. Self-reactions will form phenyl ($CH_2CCH + CH_2CCH \rightarrow C_6H_5 + H$). Phenyl reacts barrierless with atomic H to form benzene, which at the low temperatures of the ISM (excess kinetic energy of H is $\simeq 0.001$ eV compared to a binding energy of 4.5 eV) can be expected to stabilize radiatively.[187] Given the high abundance of atomic H in dense cores ($\simeq 2 \times 10^{-4}$), this reaction will then dominate over further condensation reactions.

Further growth from benzene to larger PAHs will have to involve the main hydrocarbon reservoir in molecular cloud cores. Figure 21 summarizes potential, astrophysically relevant PAH formation routes. As discussed above, the HACA mechanism only operates at high temperatures as substantial barriers are involved.[150] Vinylacetylene has been detected in TMC1[188] at an abundance of $\simeq 10^{-9}$, and the barrierless reaction of phenyl with vinylacetylene (hydrogen abstraction vinyl-acetylene addition, HAVA) forms benzene at low temperature.[182] Phenyl-Addition-DehydroCyclization (PAC) is another potential growth route toward larger PAHs at high temperatures. Radical−Radical Reactions (RRR) utilizing methyl and propargyl are also an alternative, high-temperature route.[182] The Methylidyne-Addition−Cyclization-Aromatization (MACA) route converts a vinyl side chain ($C_2H_3$) of a PAH into a five-membered ring via a ring annulation. As this figure illustrates, many reactions are barrierless and would proceed at the low temperatures of dark cloud cores. However, all require the presence of aryl radicals and H-removal from a PAH is endothermic by $\simeq 4.5$ eV. As indicated above for phenyl, the high abundance of atomic H may make these unlikely routes in dense clouds.

There may be alternative routes toward PAH formation at low temperatures. Benzene will react barrierless with small hydrocarbon radicals (e.g., ethynyl, $C_3H$) leading to phenylacetylene and diphenylacetylene.[189,190] We note that the former has been detected in TMC1. Diphenylacetylene can lead to naphthalene in an exothermic reaction stabilized by IR relaxation.[191] Alternatively, the electron affinity of PAHs larger than $\simeq 30$ C-atoms exceeds 1 eV and such PAHs will readily form anions.[36,66] Anion-molecule reactions at low temperatures will proceed at the Langevin rate. Moreover, the Coulomb interaction may potentially reduce energy barriers. As $G_0/n_H \simeq 10^{-4}$, reactions of PAHs with atomic hydrogen will lead to superhydrogenated PAHs. The chemistry of such species with small hydrocarbon radicals has not been explored yet.

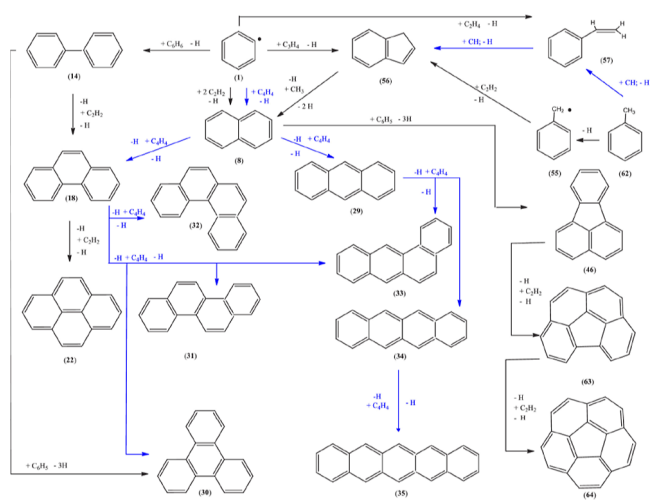

**Figure 21.** Schematic summary of potential pathways to grow aromatic species. Pathways color-coded in blue represent barrierless reactions. The multiple pathways reflect the tendency of C-rich mixtures to form aromatic systems. Considering the phenyl radical as the starting point, the first step in the HACA route is indicated by the black arrow leading from 1 to 8. Reaction with vinylacetylene (HAVA) forms an alternative barrierless route. The PAC route is indicated by the reaction of phenyl to benzene (1 to 14). The MACA route involves 57 to 8 through 56. In the RRR route, the flow is from 1 to 56, to 8. Figure taken from,[182] which should be consulted for details. Copyright [2021] American Chemical Society.

For all routes, kinetics present an underlying issue. For radical-PAH reactions, the reaction rate coefficient could be as large as $10^{-10}$ cm$^3$ s$^{-1}$ and, with an abundance of $C_2H$ and $C_4H$ of $X \simeq 10^{-7}$, the time scale for growth is $\simeq 10^4$ yr/C-atom. Other potential feedstock species are even less abundant. Models imply a time scale for the C-rich phase in the chemical evolution of the TMC1 core of only $\simeq 5 \times 10^5$ yr.[7,185] and, hence, PAHs can only grow to modest sizes. Possibly, this can be somewhat extended as atomic O will condense out slightly faster than CO due to its slightly higher thermal velocity, and CO is the ultimate source of C for hydrocarbon radical formation (through cosmic ray produced photons).

Finally, it is worth noting that the rich hydrocarbon chemistry implied by the presence of small PAHs inside dense cloud cores also has implications for the inventory of the AIB carriers in PDRs. These species will have been part of the dense core in which the massive star formed and may have grown substantially under those conditions. The presence of methyl groups and PAH clusters deep inside PDRs may well attest to this rich chemistry.

## 10. SUMMARY AND FUTURE DIRECTIONS

Over the last 40 years, we have made great strides in our understanding of the molecular Universe. Much of this was driven by, on the one hand, the development of new and improved astronomical instruments to study the molecular inventory of the Universe and, on the other hand, by chemists bringing their facilities, insights, expertise, and experience to bear on astronomical inquiries.

The launch of JWST with its exquisite spectral and spatial resolution in the mid-IR, provides almost weekly new insights in the characteristics of the AIBs and its carriers. As discussed in this review, the pattern of CH out-of-plane bending bands in the 11−14 $\mu$m range, the relatively simple pattern of CC stretching modes in the 6−9 $\mu$m range, the presence of only two AIBs in the CCC bending modes in the 15−17 $\mu$m range all attest to the predominance of highly symmetric, compact PAHs with long straight edges and few corners in the interstellar PAH family. There is a smidgen of asymmetric PAHs, of PAHs with methyl groups, of PAHs with carbonyl groups, and of deuterated PAHs. The abundance of these trace species decreases toward the surface of the PDR, likely because these groups are very susceptible to UV photolysis. In contrast to earlier suggestions, there is little evidence for the presence of N in the PAH molecular structure. There is also no evidence for cyanogen groups associated with the AIB carriers. Deep in the PDR— typically beyond the H$_2$ dissociation front—the 11.25 $\mu$m band points toward the presence of PAH clusters. Likely, the plateaus underneath the AIBs are also carried by clusters and hence, some of these clusters seem to persist to the PDR surface. Current JWST studies have focused on a few prototypical PDRs—the Orion Bar, NGC 7023, the Horsehead—and it is important to extend the JWST revolution to a broad sample of environments spanning a wide range of conditions. In this regard, early observational studies have classified AIB spectra into class $\mathcal{A}$ through $\mathcal{D}$ and sketched an evolutionary link between these classes. JWST could put the interrelationship of these classes on a firm footing and, in particular, elucidate whether the class $\mathcal{D}$ AIB spectra present in T Tauri stars bear any relationship to the normal, class $\mathcal{A}$, AIB spectra. Likewise, class $\mathcal{C}$ spectra are typically associated with a very special class of post-AGB objects (the so-called 21 $\mu$m objects). These are descendants of low mass (1 M$_\odot$), low metallicity stars and their relevance for the general interstellar PAH population is unclear. Here too, further observational study focusing on the putative evolutionary relation with class $\mathcal{A}$ spectra could be very illuminating. JWST observations of a wide sample of PDRs could also be instrumental in identifying the "end members" of the spectral components in the 6−9 $\mu$m range. The presence of only a few, primary AIBs in the CC stretching mode region is a key piece of evidence pointing toward the dominance of grandPAHs in the interstellar PAH family. As Figure 5 illustrates, the AIB pattern in this wavelength range can be considered a fingerprint region; particularly when considering the subclass of highly symmetric and compact PAH cations in the size range of 30−90 C-atoms. A future far-IR spectroscopy space mission combined with JWST can provide complete 3−300 $\mu$m AIB spectra of a large sample of source. The far-IR vibrational modes of PAHs—i.e., the jumping jack and drumhead modes—involve the C skeleton and are very characteristic for individual molecules and this might provide the best identification route of the AIB carriers.

Dedicated efforts in laboratory spectroscopy and quantum chemistry have established that anharmonic DFT calculations provide frequencies with an accuracy $\simeq 0.2\%$. Further improvements will require more accurate basis sets and functionals, the inclusion of higher order terms in the description of the potential energy surface, and the inclusion of three (or more) combination bands in the resonances. Current studies are limited to relatively small species, $\lesssim 40$ C-atoms. As computer costs increase with the size of the species to the fourth power, progress in this area may be slow. As current studies show, anharmonicity plays a key role in spectroscopy of small PAHs and possibly the way forward will be to identify clear trends in the existing spectroscopic data and use this to extrapolate our knowledge to large PAHs. Also, the emphasis has been on neutral PAHs and the effects of anharmonicity in the spectra of PAH cations are largely unexplored.

The emission process is a cascade in energy from an initial high energy. Anharmonicity linked to spectator modes leads to peak shifts, broadening, and red-shaded profiles. These effects are well modeled for small neutral PAHs. Generalizing these results to large PAHs will require a good grasp of the size-dependence of anharmonicity. As these effects are highly dependent on the internal energy of the emitting species, a deep understanding of this may provide an independent handle on the size of the carriers of the AIBs. Observations of the interstellar $C_{60}$ bands may provide a clear test of these aspects as—rather than adopting extrapolated/generic properties—the molecular properties of $C_{60}$ are well-defined and amenable to laboratory and quantum chemical studies. Clearly, a concerted, observational, astronomical modeling, experimental, and quantum chemistry focus on $C_{60}$ would be very valuable.

Recent studies have highlighted the importance of recurrent electronic fluorescence as a relaxation process for small PAH cations. This process may skew the AIB spectra toward a subclass of the interstellar PAH family. It will also have direct consequences for the analysis and interpretation of the AIBs in terms of abundance, size, and degree of ionization. Further studies of large neutral and cationic PAHs would be very valuable.

Experimental and modeling studies have established that PAH photochemistry is a competition between fragmentation and isomerization. The fragmentation process is dominated by loss of the weakest link: breakup will start with cluster dissolution, loss of "super-H", $CH_3$ loss, and then loss of aromatic H. For compact PAHs, loss of $C_2H_2$ requires very high internal energies or high levels of dehydrogenation. In that sense, aromatic H acts as a safety valve for the breakup of the C-skeleton of interstellar PAHs as the reverse reactions of aryl radicals with atomic H is fast. While observations reveal the loss of $CH_3$ groups and clusters in PDRs, fragmentation of large PAHs is limited to the UV-rich surface layers. Characterizing the photochemical evolution of (large) PAHs in PDR surfaces and identifying potential grandPAHs may establish whether the interstellar PAH population is a "nuclear" family or a large and dispersed group. As interstellar $C_{60}$ may be the photochemical grandchild of large PAHs, linking the $C_{60}$ abundance to the physical conditions in PDR will help addressing these issues.

The detection of a wide range of small aromatic species inside dense molecular cloud cores is one of the exciting recent developments in the field. These studies herald the presence of active gas phase routes toward small aromatic molecules under these shielded conditions and they carry the promise of deep insights in the chemical routes involved. This has initiated a rich array of focused experimental/quantum chemistry studies on potentially astrochemically relevant reactions. It might be good to extend such studies to PAH anions and to superhydrogenated PAHs as both are expected to be abundant in dense cloud cores. We can expect that similar reaction routes will be open to large PAHs as well. In that respect, high abundances of $C_2H$ are present in the hydrocarbon radical zone in the Orion Bar surface and further combined ALMA/JWST studies of the AIB characteristics and $C_2H$ abundances may address PAH growth under these conditions.

The identification of 5 DIBs with electronic transitions in $C_{60}^+$ has given new impetus to observational and experimental studies in this field. Fullerenes, smaller carbon cages, PAHs and their derivatives are obvious candidates for these absorption features. The HOMO−LUMO band gap for neutral PAHs is > 2 eV for neutral compact PAHs with less than 80 C-atoms and they are not good candidate DIB carriers. However, the HOMO−LUMO band gap for the neutral acene or rylene families is already less than 2 eV for 20−35 C-atoms and that makes this class interesting DIB candidates. Electronic transitions of PAH cations will occur at much lower energies than their neutral parents and spectra of such species have been a focus in experimental investigations. But as a counterargument, it should be kept in mind that PAHs are expected to be mostly neutral in diffuse clouds, making PAH cations perhaps unlikely candidates both because of reasons of required absolute abundances and expected large abundance variations between sightlines as the physical conditions can vary substantially in diffuse medium sightlines.

With $\simeq 500$ DIBs detected, identification of individual DIBs with specific molecules is daunting as there likely is a (very, very weak) DIB at any wavelengths when integrated deep enough. Past studies of DIB carriers have focused on species that were thought to be abundant because of thermodynamic stability arguments—a strategy that has proven to be very successful for $C_{60}^+$. However, interstellar chemistry is driven by kinetics rather than thermodynamics and, for example, photodriven isomerization may play an important role. Taking the observed AIB characteristics in PDR spectra as a guide may also be fraught with issues as PAHs will be heavily "cooked" in PDR surfaces and differ considerably from the PAH family in the diffuse ISM. It might be interesting to link the inventory of complex hydrocarbons in dense cloud cores with the DIB characteristics of sightlines traversing the translucent boundaries of these clouds. Another potential avenue might be to analyze the variations in the AIB characteristics of the Red Rectangle with JWST and link those to the variations in the molecular emission bands in the visible. This has the additional advantage that it may be easier to link vibronic structure in the electronic transitions to the vibrational structure as revealed by the AIBs. Moreover, this will automatically focus the study on the brightest DIBs, which are the most interesting ones anyway. The importance of recurrent fluorescence indicated by the visible emission bands is another interesting avenue of research to follow.

## ■ AUTHOR INFORMATION

**Corresponding Author**

**Alexander G. G. M. Tielens** − *Department of Astronomy, University of Maryland, College Park, Maryland 20742-2421, United States;* ● orcid.org/0000-0003-0306-0028; Email: tielens@umd.edu

## Notes

The author declares no competing financial interest.

## ACKNOWLEDGMENTS

I gratefully acknowledge the many students, postdoctoral fellows and collaborators who have enlightened me with their insights in the characteristics of interstellar PAHs and their role in the molecular Universe.

## ADDITIONAL NOTES

$^a$The spectral ranges of the PAH OOPs modes have been taken from the PAHdb: https://www.astrochemistry.org/pahdb/theoretical/3.20/default/view[19].

$^b$Edge structures with two bound carbon atom that are not bound to any hydrogen atoms.

$^c$Symmetry point group is the classification of the symmetry of a molecule where a set of symmetry operations leaves at least one point unperturbed. $C_1$, only symmetry group is $360°$ rotation; $C_s$, single plane of symmetry; $C_{2h}$, a 2-fold rotation axis, a horizontal mirror plane and a center of inversion.

$^d$Class $C$ and $\mathcal{D}$ are more peculiar objects, such as metal-poor, post-AGB objects descending from $1\ M_\odot$ progenitors.

$^e$PAHs with one or more nitrogen atoms substituted into their carbon skeleton.

$^f$Bath modes that are not involved in the transition.

$^g$The only obvious difference is in the weak 11.25 $\mu$m band on the shoulder of the 11.2 $\mu$m AIB. In the Orion Bar, this feature, due to PAH clusters, is prominent in the region beyond the H$_2$ dissociation front, DF3[18].

$^h$acenes refer to the sequence napthalene, anthracene, tetracene, ...Rylenes refer to the sequence perylene, terrylene, quarterrylene, $\cdots$

$^i$GOTHAM: GBT Observations of TMC-1: Hunting Aromatic Molecules program; & QUIJOTE: Q-band Ultrasensitive Inspection Journey to the Obscure TMC-1 Environment program.

$^j$So-called Franck–Condon Herzberg–Teller interference.

$^k$Actually, phenyl (C$_6$H$_5$).

$^l$Photon energy, $E_{ap}$, where products first appear at $\simeq$ 10% level; i.e., where the fragmentation rate equals $\simeq$ 10% of the inverse of the experimental time scale ($k(E_{ap}) \simeq 10^4$ s$^{-1}$).

$^m$Magic numbers are clusters that show enhanced stability compared to neighboring clusters as revealed by their higher abundance in formation/destruction experiments.